\documentstyle[vsolj01,graphicx,natbib,amsmath,amssymb]{article}

\def\cite{\citealt}

\begin{document}

\title{Optical Variability Correlated with X-ray Spectral Transition}
\title{in the Black-Hole Transient ASASSN-18ey = MAXI J1820+070}

\author{Keito Niijima$^{1}$, Mariko Kimura$^{1}$, Yasuyuki Wakamatsu$^{1}$, Taichi Kato$^{1}$, Daisaku Nogami$^{1}$, Keisuke Isogai$^{1}$,} \author{Naoto Kojiguchi$^{1}$, Ryuhei Ohnishi$^{1}$, Megumi Shidatsu$^{2}$, Geoffrey Stone$^{3}$, Franz-Josef Hambsch$^{4,5,6}$,} \author{Tam\'as Tordai$^{7}$, Michael Richmond$^{8}$, Tonny Vanmunster$^{9,10}$, Gordon Myers$^{11}$,
Stephen M. Brincat$^{12}$,} \author{Pavol A. Dubovsky$^{13}$, Tomas Medulka$^{13}$, Igor Kudzej$^{13}$, Stefan Parimucha$^{14}$, Colin Littlefield$^{15}$,} \author{Berto Monard$^{16,17}$, Joseph Ulowetz$^{18}$, Elena P. Pavlenko$^{19}$, Oksana I. Antonyuk$^{19}$, Aleksei A. Sosnovskij$^{19}$,} \author{Aleksei V. Baklanov$^{19}$, Kirill A. Antoniuk$^{19}$, Nikolai V. Pit$^{19}$, Sergei P. Belan$^{19}$, Julia V. Babina$^{19}$,} \author{Aleksandr S. Sklyanov$^{20}$, Anna M. Zaostrozhnykh$^{21}$, Andrew V. Simon$^{22}$, Lewis M. Cook$^{23}$, Ian Miller$^{24}$,} \author{Hiroshi Itoh$^{25}$, Domenico Licchelli$^{26,27}$, Shawn Dvorak$^{28}$, Richard Sabo$^{29}$, Yenal \"{O}gmen$^{30}$, Donn R. Starkey$^{31}$,} \author{Peter Nelson$^{32}$, Enrique de Miguel$^{33,34}$, Charles Galdies$^{35}$, Kenneth Menzies$^{36}$, Seiichiro Kiyota$^{37}$,} \author{Arto Oksanen$^{38}$, Roger D. Pickard$^{39,40}$, Alexandra M. Zubareva$^{41,42}$, Klaus Wenzel$^{5}$ \& Denis Denisenko$^{42}$}
\author{}

\author{$^1$ Department of Astronomy, Graduate School of Science, Kyoto University,}
\author{Oiwakecho, Kitashirakawa, Sakyo-ku, Kyoto 606-8502, Japan}
\author{$^2$ Department of Physics, Ehime University, 2-5, Bunkyocho, Matsuyama, Ehime 790-8577, Japan}
\author{$^3$ American Association of Variable Star Observers, 49 Bay State Rd., Cambridge, MA 02138, USA}
\author{$^4$ Groupe Europ\'een d’Observations Stellaires (GEOS), 23 Parc de Levesville, 28300 Bailleau l’Ev\^eque, France}
\author{$^5$ Bundesdeutsche Arbeitsgemeinschaft f\"{u}r Ver\"{a}nderliche Sterne (BAV), Munsterdamm 90, 12169 Berlin, Germany}
\author{$^6$ Vereniging Voor Sterrenkunde (VVS), Oude Bleken 12, 2400 Mol, Belgium}
\author{$^7$ Polaris Observatory, Hungarian Astronomical Association, Laborc utca 2/c, 1037 Budapest, Hungary}
\author{$^8$ Physics Department, Rochester Institute of Technology, Rochester, New York 14623, USA}
\author{$^9$ Center for Backyard Astrophysics Belgium, Walhostraat 1A, B-3401 Landen, Belgium}
\author{$^{10}$ Center for Backyard Astrophysics Extremadura, 06340 Fregenal de la Sierra, Spain}
\author{$^{11}$ Center for Backyard Astrophysics San Mateo, 5 inverness Way, Hillsborough, CA 94010, USA}
\author{$^{12}$ Flarestar Observatory, Fl.5/B, George Tayar Street, San Gwann SGN 3160, Malta}
\author{$^{13}$ Vihorlat Observatory, Mierova 4, 06601 Humenne, Slovakia}
\author{$^{14}$ Institute of Physics, Faculty of Science, UPJS Kosice, Slovakia}
\author{$^{15}$ Department of Physics, University of Notre Dame, } \author{ 225 Nieuwland Science Hall, Notre Dame, Indiana 46556, USA}
\author{$^{16}$ Bronberg Observatory, Center for Backyard Astrophysics Pretoria,} \author{PO Box 11426, Tiegerpoort 0056, South Africa}
\author{$^{17}$ Kleinkaroo Observatory, Center for Backyard Astrophysics Kleinkaroo,} \author{ Sint Helena 1B, PO Box 281, Calitzdorp 6660, South Africa}
\author{$^{18}$ Center for Backyard Astrophysics Illinois, Northbrook Meadow Observatory,} \author{ 855 Fair Ln, Northbrook, Illinois 60062, USA}
\author{$^{19}$ Federal State Budget Scientific Institution “Crimean Astrophysical Observatory of RAS”,} \author{ Nauchny, 298409, Republic of Crimea}
\author{$^{20}$ Kazan (Volga region) Federal University, Kremlevskaya str., 18, Kazan, 420008, Russia}
\author{$^{21}$ Institute of Physics, Kazan Federal University, Ulitsa Kremlevskaya 16a, Kazan 420008, Russia}
\author{$^{22}$ Astronomy and Space Physics Department, Taras Shevshenko National University of Kyiv,} \author{ Volodymyrska str. 60, Kyiv, 01601, Ukraine}
\author{$^{23}$ Center for Backyard Astrophysics Concord, 1730 Helix Ct. Concord, California 94518, USA}
\author{$^{24}$ Furzehill House, Ilston, Swansea, SA2 7LE, UK}
\author{$^{25}$ Variable Star Observers League in Japan (VSOLJ), 1001-105 Nishiterakata, Hachioji, Tokyo 192-0153, Japan}
\author{$^{26}$ R.P. Feynman Observatory, Gagliano del Capo, 73034, Italy}
\author{$^{27}$ CBA, Center for Backyard Astrophysics - Gagliano del Capo, 73034, Italy}
\author{$^{28}$ Rolling Hills Observatory, 1643 Nightfall Drive, Clermont, Florida 34711, USA}
\author{$^{29}$ 2336 Trailcrest Dr., Bozeman, Montana 59718, USA}
\author{$^{30}$ Green Island Observatory, Ge\c{c}itkale, Magosa, via Mersin, North Cyprus}
\author{$^{31}$ DeKalb Observatory, H63, 2507 County Road 60, Auburn, IN 46706, USA}
\author{$^{32}$ 1105 Hazeldean Rd, Ellinbank 3820, Australia}
\author{$^{33}$ Departamento de Ciencias Integradas, Facultad de Ciencias Experimentales,} \author{ Universidad de Huelva, 21071 Huelva, Spain}
\author{$^{34}$ Center for Backyard Astrophysics, Observatorio del CIECEM, Parque Dunar,} \author{ Matalasca\~{n}as, 21760 Almonte, Huelva, Spain}
\author{$^{35}$ Institute of Earth Systems, University of Malta}
\author{$^{36}$ Center for Backyard Astrophysics (Framingham), 318A Potter Road, Framingham, MA 01701, USA}
\author{$^{37}$ VSOLJ, 7-1 Kitahatsutomi, Kamagaya, Chiba 273-0126, Japan}
\author{$^{38}$ Hankasalmi observatory, Jyvaskylan Sirius ry, Verkkoniementie 30, FI-40950 Muurame, Finlandv}
\author{$^{39}$ The British Astronomical Association, Variable Star Section (BAA VSS),} \author{ Burlington House, Piccadilly, London, W1J 0DU, UK}
\author{$^{40}$ 3 The Birches, Shobdon, Leominster, Herefordshire, HR6 9NG, UK}
\author{$^{41}$ Institute of Astronomy, Russian Academy of Sciences, Moscow 119017, Russia}
\author{$^{42}$ Sternberg Astronomical Institute, Lomonosov Moscow State University,} \author{ Universitetsky Ave., 13, Moscow 119992, Russia}

\begin{abstract}
\textbf{ How a black hole accretes matter and how this process is
regulated are fundamental but unsolved questions in astrophysics.  
In transient black-hole binaries, a lot of mass stored 
in an accretion disk is suddenly drained to the central 
black hole because of thermal-viscous instability.  This phenomenon 
is called an outburst and is observable at various 
wavelengths \citep{AccretionPower3}.  
During the outburst, the accretion structure in the vicinity of 
a black hole shows dramatical transitions from a geometrically-thick 
hot accretion flow to a geometrically-thin disk, and 
the transition is observed at X-ray 
wavelengths \citep{rem06BHBreview,don07XB}.  
However, how that X-ray transition occurs remains a major 
unsolved problem \citep{dun08gx339}.  
Here we report extensive optical photometry during the 2018 
outburst of ASASSN-18ey (MAXI J1820$+$070), a black-hole 
binary at a distance of 3.06 kpc \citep{tuc18a18ey,tor19j1820}
containing a black hole and a donor star of less than 
one solar mass.
We found optical large-amplitude periodic variations
similar to superhumps which are well observed 
in a subclass of white-dwarf binaries \citep{Pdot}.
In addition, the start of the stage transition 
of the optical variations was observed 5 days earlier than 
the X-ray transition.  This is naturally explained
on the basis of our knowledge regarding white dwarf 
binaries as follows:
propagation of the eccentricity inward in the disk makes 
an increase of the accretion rate in the outer disk, 
resulting in huge mass accretion to the black hole.
Moreover, we provide the dynamical estimate of the binary 
mass ratio by using the optical periodic variations
for the first time in transient black-hole binaries.
This paper opens a new window to measure black-hole masses
accurately by systematic optical time-series observations which
can be performed even by amateur observers. }

\end{abstract}

On March 6th, 2018, ASASSN-18ey was discovered 
by the All-Sky Automated Survey for SuperNovae (ASAS-SN) 
with an increase of optical luminosity. This object was 
faint about 4 days before the discovery, and hence, this 
sudden optical eruption was the onset of an outburst.  
After the detection, on March 12th, 2018, the Monitor of 
All-sky X-ray Image (\textit{MAXI}) found an X-ray transient 
MAXI J1820$+$070 \citep{kaw18atel11399} which was identified 
with ASASSN-18ey via optical follow-up 
observations \citep{den18atel11400}. 
This object was suggested to be a black-hole binary 
from the X-ray spectra and the typical relation 
between optical and X-ray luminosity \citep{bag18atel11418}.  
Also, very recent spectroscopic observations 
confirmed that the central object in this binary is 
a black hole by estimating the binary mass function \citep{tor19j1820}. 

We performed multi-color optical photometry of
ASASSN-18ey = MAXI J1820$+$070 via the Variable 
Star Network (VSNET) since BJD 2,458,189 (corresponds to 2018 March 11)
(Extended Data 
Table 1 and Methods section `Detailed methods of optical 
observations and analyses').  
We also used the optical data taken by the American Association 
of Variable Star Observers (AAVSO).  
Our data are one of the densest optical data of 
outbursts in transient black-hole binaries in the sense that 
they contain $>$400,000 data points and cover the entire 
outburst that lasted for more than 200 days.  
We also compare our optical data with the 2.0--4.0-keV and 
4.0--10.0-keV X-ray light curves taken by \textit{MAXI}.   
   
The overall optical and X-ray light curves of this outburst 
are shown in Figure 1a.  The rapid rise and the exponential 
decay of the optical behavior seems to be typical during 
outburst in black-hole binaries \citep{tan96XNreview}.  
We did not detect any optical periodic variations during 
the first 70 days.  
However, large-amplitude periodic oscillations appeared 
around BJD 2,458,260 (corresponds to 2018 March 21)
and continued for $\sim$145 days 
(Figure 1b).  
Here we use the two-dimensional least absolute shrinkage 
and selection operator (Lasso \citet{kat12perlasso}) 
to derive sharp peaks in power spectra from unevenly 
sampled data by suppressing aliases.
Although the light curves of soft-X-ray outbursts in typical 
black-hole binaries show sudden increases at the initial 
stage and long plateaus lasting a few hundred
days \citep{rem06BHBreview,don07XB}, 
the initial increase of soft-X-ray flux was moderate 
in this object, and the dramatic X-ray spectral transition 
began $\sim$100 days after the onset of the X-ray outburst.  

\begin{figure}[htbp]
	\centering
           \includegraphics[width=150mm]{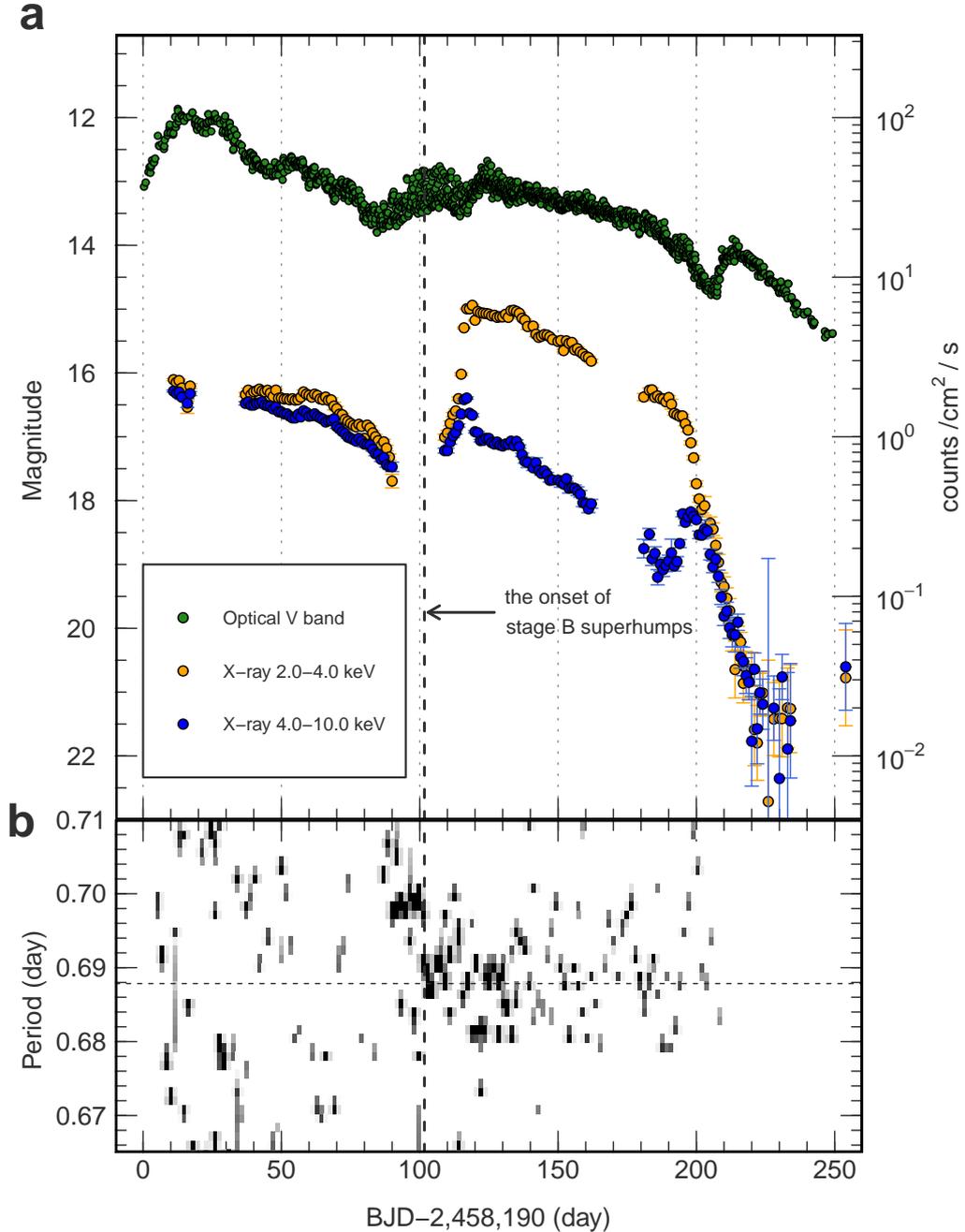}
	\caption{\textbf{Overall light curves of ASASSN-18ey and the two-dimensional power spectral density by Lasso.} Panel (a) shows optical $V$-band and MAXI X-ray light curves of ASASSN-18ey during Barycentric Julian Day (BJD) 2,458,189--2,458,440. The horizontal and vertical axis represents BJD-2,458,190 and Magnitude respectively. The vertical dashed line represents the onset of stage B superhumps. There seems to be a relation between the X-ray state transition and the change of superhump stage. Error bars in the X-ray light curves represent 1$\sigma$ confidence intervals. The errors in optical light curve are equal or smaller than the size of the points. Panel (b) shows the time-varying periods calculated by Lasso which can derive a sharp peak from a sparse light curve. The vertical axis represents the period. Once the superhumps started, the characteristic period appeared.
        }
        \label{lc}
\end{figure}

We investigated the period and amplitude variations of 
the detected optical periodic variability in ASASSN-18ey 
(Figure 2).
Their periods were always longer than the detected orbital 
period 0.68549(1)~days, which was estimated 
from the radial velocities of the binary 
motion \citep{tor19j1820}.  
The $O-C$ plane shown in Figure 2a represents 
the accumulated time difference of periods 
with respect to the reference period at each cycle E and 
is sensitive to slight period variations (Extended Data 
Table 2 and Methods section `SU UMa/WZ Sge-type stars and 
time evolution of their superhumps').  
We found that the period was constant and that the amplitude 
increased during around BJD 2,458,260--2,458,290.  After that, 
the period suddenly dropped, and gradually increased, 
while the amplitude gradually decreased.
The positive excesses of the periods with respect to 
the orbital period and other characteristic features are 
identical to ``superhumps'' observed during outburst in 
SU UMa-type stars, a subclass of dwarf novae with small 
mass ratios of the donor to the primary \citep{Pdot}.  
Dwarf novae are close binary systems consisting of
a central white dwarf and a donor star \citep{war95book}.
Superhumps have periods slightly longer than
the corresponding orbital period, and the periods 
vary with time.  
They are believed to be generated by a periodic tidal 
dissipation in the precessing eccentric disk which is produced
by tidal instability working in binaries having  
mass ratios smaller than 0.25 (Methods section 
`SU UMa/WZ Sge-type stars and time evolution of their 
superhumps').  Hereafter, we call the constant-period stage 
``stage A'' and the varying-period stage ``stage B'', 
following the notation in the field of dwarf novae 
(Methods section `SU UMa/WZ Sge-type stars and time evolution 
of their superhumps' ).

\begin{figure}[htbp]
	\centering
		\centering\includegraphics[width=100mm]{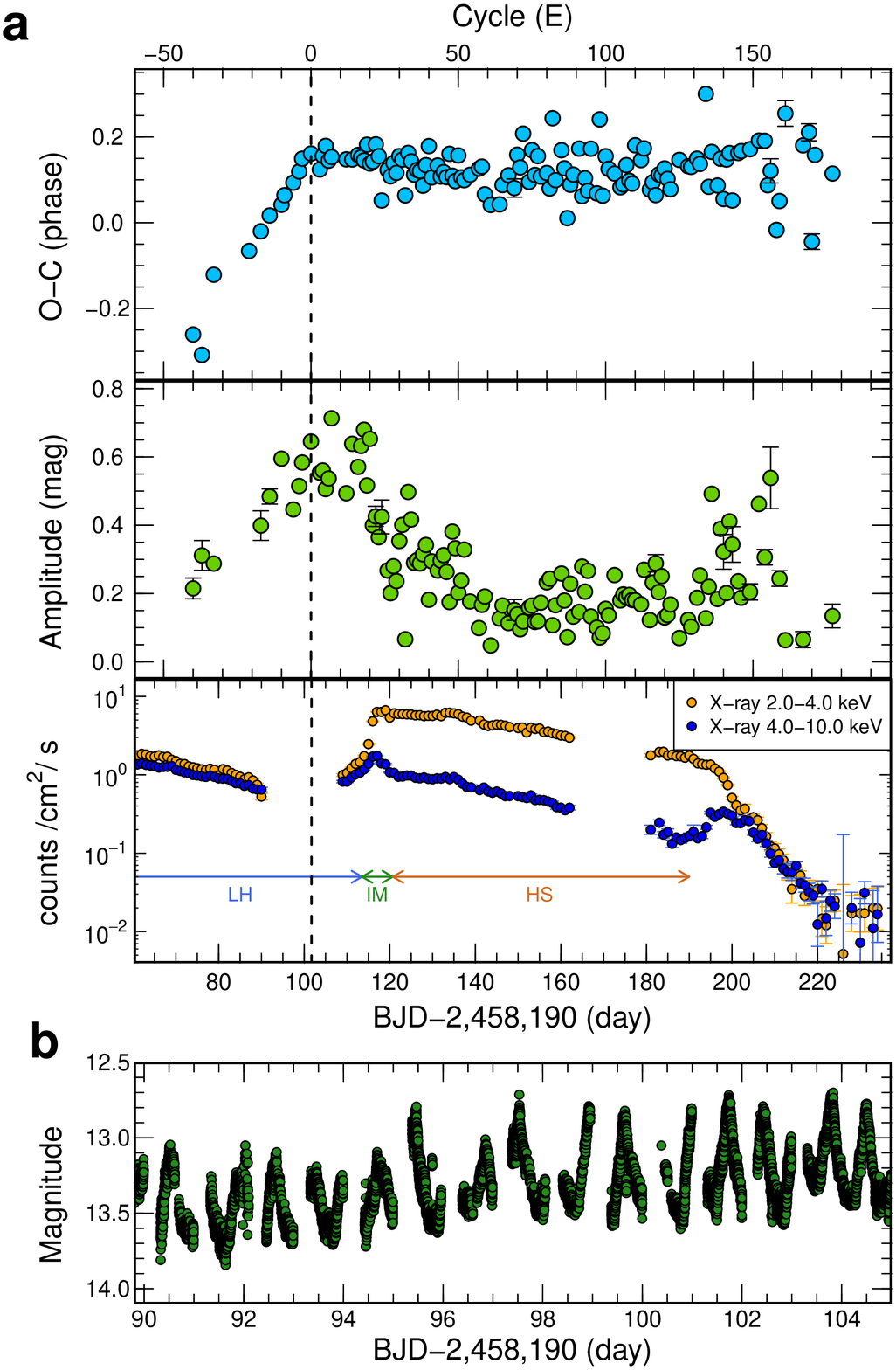}
	\caption{\textbf{The comparison of the stage of superhumps and X-ray state transition in ASASSN-18ey.}  (a)Upper: The $O-C$ of ASASSN-18ey, which expresses the time variation of the period of superhumps. Middle: The variation of the amplitude of superhumps. Bottom: The X-ray light curves of ASASSN-18ey. In these panels, the error bars represent 1$\sigma$ confidence intervals. On BJD 2,458,291.7 denoted by the vertical dashed line, there is a clear bending point on the $O-C$ and the amplitude planes, which means that the X-ray hard-to-soft state transition is a slightly delayed with respect to the stage transition of superhumps. The X-ray spectral state reported by \citet{shi19j1820} is also denoted in the bottom panel and ``LH'', ``IM'', and ``HS'' stand for the low-hard state, the intermediate (very high state), and the high-soft state, respectively. We here use C = 2458261.952323$+$0.688907 E. (b)The light curve of superhumps during stage A. The single-peaked humps are clearly observed. The amplitude is $\sim$0.8 mag and the averaged period is $\sim$0.697 days.}
	\label{oc}
\end{figure}

As stated above, dwarf novae have accretion disks 
just like black-hole binaries.  Since these objects have
the different scale and temperature but have the similar
structure, dwarf novae are sometimes called ``a miniature
version of black-hole binaries'' \citep{kuu96TOAD}. 
Taking this similarity into consideration,
it is expected that the physics of an accretion disk of
a black-hole binary may be understood by analogy to 
the underlying physics of superhumps in dwarf novae.

Although superhump candidates have been reported only in 
several black-hole binaries prior to this work, 
nobody found clearly their period variations (Methods 
section `Black-hole binaries showing superhump candidates 
in the past'). 
Our study establishes the presence of 
superhumps as a process that may affect black-hole X-ray 
binaries that undergo canonical evolution through multiple 
accretion states during an outburst.  
Although the averaged amplitudes and period of superhumps 
in SU UMa stars are typically $\sim$0.25 mag and 
less than 0.1 days, respectively \citep{Pdot3}, those of 
ASASSN-18ey are much larger ($\sim$0.8 mag) and longer 
($\sim$0.69 days).
The long period of superhumps is naturally 
expected from the orbital period of this system \citep{tor19j1820}, 
which is much longer than typical orbital periods 
in SU UMa-type dwarf novae.
On the other hand, we predict the superhump amplitude 
should be $\sim$0.07 mag in this system if the intrinsic 
superhump amplitude is the same as the averaged amplitude 
in SU UMa stars and the greater X-ray irradiation of 
ASASSN-18ey is accounted for. 
The superhumps caused by the tidal dissipation are thus 
detectable against previous suggestions \citep{has01BHXNSH}.
However, the expected amplitude is smaller than 
the observational one.
The $\sim$100-times larger disk than the averaged one 
in SU UMa stars and/or the change of the surface area 
receiving X-rays from the inner disk might contribute to
the large amplitude of the superhumps (Figure 7 and 
Methods section `SED modelling').  

We estimated the period of stage A superhumps to be 0.7029(3)~days 
by using the data during BJD 2,458,263.7--2,458,291.7
(corresponds to 2018 May 25.2--2018 June 22.2).  
Then we were able to estimate the binary mass ratio ($q$)
of this object to be 0.066(1) from the stage A superhump
period and the identified orbital period \citep{tor19j1820}
by the dynamical method for mass-ratio estimations, which is 
based on a well-established treatment of celestial mechanics and
is verified in the study of dwarf novae \citep{kat13qfromstageA}
(Methods section 'Estimation of black-hole masses with 
stage A superhumps'). 
The old method proposed by \citet{pat05SH} is not suitable, 
since it is based on the empirical relation between 
the orbital period and the period of time-varying stage B 
superhumps.  It was proved that this method easily underestimate 
the binary mass ratio \citep{nak13j2112j2037}.
Our estimate is more accurate than ever because we used 
stage A superhumps only.
Since the spectral type of the donor is constrained to 
be K late type, the donor mass is
$0.59^{+0.11}_{-0.10}$ $M_{\odot}$ \citep{tor20j1820}, where 
$M_{\odot}$ is the solar mass.  
Therefore the black-hole mass is estimated to be 
7.3--10.8$M_{\odot}$.  This is consistent with the black-hole mass 
estimated from the rotational velocity to be 
8.48$^{+0.79}_{-0.72}$~$M_{\odot}$ \citep{tor20j1820}. 

Also, we obtain 0.047--0.093 as the value of the mass ratio, 
which infers the black-hole mass to be 5.3--12.2$M_{\odot}$,  
by using a relation between the duration of 
stage A superhumps and the mass ratio (Figure 3; Methods 
section 'Estimation of black-hole masses with stage A 
superhumps').  
This relation is based on the theory \citep{lub91SHb}
that the growth time of superhumps is proportional to $q^{-2}$, 
and has also been confirmed observationally (Methods section 
`Estimation of black-hole masses with stage A superhumps').
The duration of the stage A superhumps in ASASSN-18ey is 
estimated to have been 71 cycles if the epoch at which 
the superhump amplitude should have been zero is BJD 2,458,238.5 (corresponds to 2018 April 30.0).  
We have derived this epoch by the linear regression of the data 
before BJD 2,458,291.7 which is displayed in the second panel of 
Figure 2a and have calculated the cycle length by dividing 
the estimated stage A superhump duration by its period.  
We have substituted this estimate into the relation 
denoted by the black solid line in Figure 3, which is 
derived by the linear regression of the orange circles in 
the same figure.  We consider the error of the regression 
to obtain the mass ratio of ASASSN-18ey.
The estimated range of the black-hole mass covers that derived 
by the former method. 

\begin{figure}[htbp]
	\centering
	  \centering\includegraphics[width=80mm]{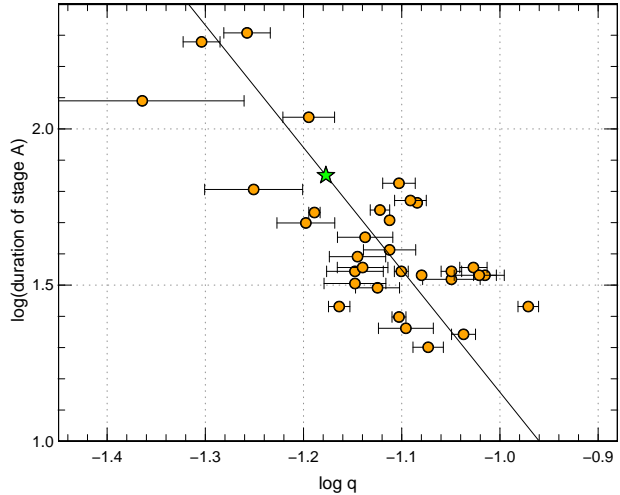}
        \vskip15mm
	\caption{\textbf{The estimations of the binary mass ratio by stage A superhumps.} The relation between the duration of stage A and the mass ratio $q$. The vertical axis is the duration of stage A (cycle) in log scale and the horizontal axis is mass ratio in log scale. The orange points represent WZ Sge-type objects. The oblique line is the result of a linear regression analysis. Using this relation, ASASSN-18ey is located at the green point from its duration of stage A. Its mass ratio, therefore, can be obtained and the value is 0.065--0.071. This error comes from the indeterminateness of the duration of stage A (we postulated that the duration is 50--60 cycles). The error bars represent 1$\sigma$ confidence intervals.
        }
	\label{phave}
\end{figure}

The inclination angle is estimated to be $\sim$60 deg from
our estimates of the binary mass ratio,
if we assume the mass function of ASASSN-18ey is 
$\sim$5.18 \citep{tor19j1820}.  This result is consistent
with no confirmed eclipsing events in this system and 
the estimated orientation of the radio jet \citep{art20j1820}, 
although the inferred inclination by \citep{tor20j1820} 
is arrowed to exceed 66 deg.
Torres et al.(2019) assumed that ASASSN-18ey shows the grazing eclipse,
and constrained the inclination;however,
there are no signs of that eclipse in our extensive photometry.
The inclination angle should be smaller than their estimate as
we have derived above. Also, they used the old method for estimating
the binary mass ratio. This is the reason why our estimate of the black-hole
mass is deviated from their result.
The recent remarkable development of time-domain astronomy will
provide plenty of optical light curves during bright outbursts,
and hence, the black-hole-mass estimations in transient sources
by the two methods using stage A superhumps will be rapidly spread.  

Interestingly, the stage A to B transition of superhumps 
occurred nearly at the same time as the beginning of 
the hard-to-soft state transition of X-ray spectra 
(Figure 2a).  
From these panels that the rapid increase of soft X-ray flux 
around BJD 2,458,300 (corresponds to 2018 June 30) seems to have been almost coincident 
with the onset of stage B superhumps but delayed for 
$\sim$10 days. 
The estimated disk radius during stage B superhumps 
by broadband SED analyses is $\sim$10$^{5}$~$r_{\rm g}$ 
(Methods section `SED modelling'), and the delay timescale is 
consistent with the viscous timescale at that 
radius \citep{mey84ADtransitionwave}.  
Here, $r_{\rm g}$ indicates the gravitational radius 
defined as $GM_1/c^2$.  
Before the X-ray state transition, 
an optically-thick hot accretion flow \citep{nar94ADAF,esi97gumusADAF}
producing hard photons was located above the inner disk and/or
inside a truncated disk and governed the X-ray 
spectra \citep{shi18j1820,kar19j1820,shi19j1820}.  
However, our SED analyses and other works suggest 
the inner disk outshone the hot corona after the transition 
during stage B superhumps (Figure 7 and Methods 
section `SED modelling').
We interpret that the mass accretion rate suddenly rose 
by a factor of $\gtrsim$ 2 after the state transition under 
the assumption of almost constant radiative efficiency \citep{shi19j1820}.
The X-ray delay at the onset of this outburst also had a
similar timescale, and is regarded as the viscous 
timescale of the large disk.  

We suggest that the tidal instability would contribute to 
the sudden appearance of a luminous innermost disk 
at the dramatic hard-to-soft X-ray state transition.
The observed superhumps, our mass-ratio estimation, 
and the optical behavior of the overall outburst 
strongly suggest that the tidal instability worked 
during the 2018 outburst of ASASSN-18ey in addition 
to the thermal-viscous instability
(Methods section `Overall optical behavior').
The tidal instability is believed to 
cause not only the disk precession but also the increase 
in mass accretion rates \citep{osa89suuma}.
We show in Figure 4 the schematic picture of the time 
evolution of the accretion disk when the tidal 
instability works.
Once the tidal instability is triggered, the non-axisymmetric
perturbation is amplified.
Eccentricity gradually develops only at the outermost ring 
of a disk, and the periodic tidal dissipation by the donor 
in the prograde precessing disk causes stage A superhumps.
Once the precessing ring is formed, the removal of 
the angular momentum becomes highly efficient 
at the outermost ring and the mass accretion is 
locally enhanced.
The increase in the mass accretion rate then propagates 
inwards and causes a global increase in mass accretion rates 
after a viscous time simultaneously with the development of 
stage B superhumps regarded as the onset of the inward 
propagation of the eccentricity \citep{kat13qfromstageA}.  
We interpret that this process occurred in the 2018 outburst 
of ASASSN-18ey; the accretion rate at the outer disk was 
firstly raised by the efficient angular-momentum removal 
and it propagated to the innermost region of the disk 
over $\sim$10 days, which induced the dramatic 
increase of soft-X-ray flux. 

\begin{figure}[htbp]
	\centering
		\centering\includegraphics[width=120mm]{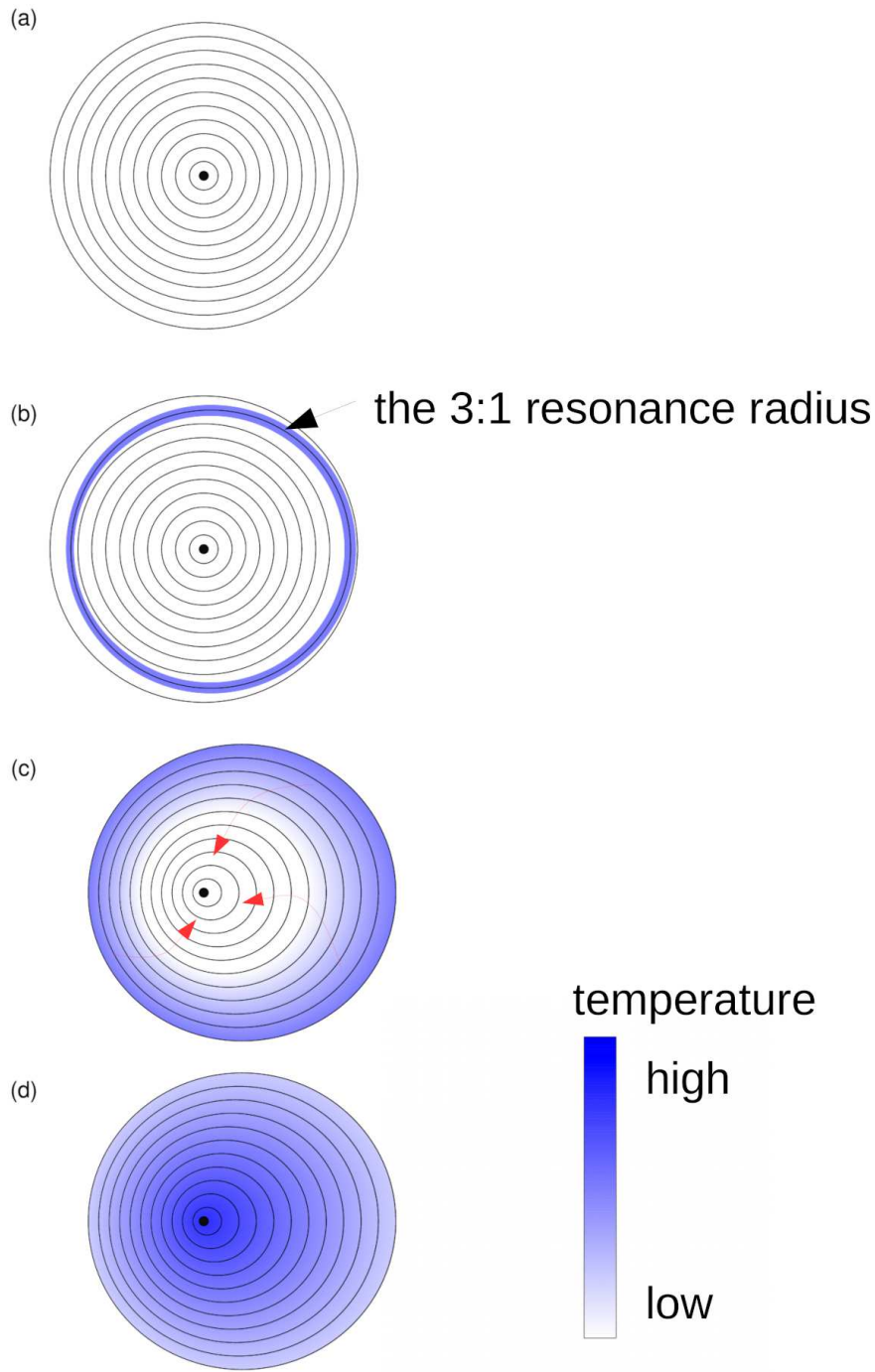}
	        \caption{\textbf{The evolution of accretion correlated with the time-varying superhump behavior.} The schematic picture of the time evolution of accretion in a disk, which is expected to be similar to that in dwarf novae. In each picture from (a) to (d), the central black points represent white dwarfs. The darker the color is, the higher the temperature is. (a) The temperature is low at the outer rim, and smoothly increases inwards in a steady-state disk. (b) The disk matter at the 3:1 resonance radius, which rotates around the white dwarf three times while the donor revolves once, is affected by a tidal force from the donor. When the stage A superhumps develop, the ring around the 3:1 resonance radius suffers from the enhanced tidal torque, and becomes eccentric. Every time the donor arrives at the periastron of the elliptical disk, the angular momentum loss at the outer disk is enhanced and the accretion rate becomes higher locally. (c) At the end of stage A superhumps, the local variations of mass accretion rates propagate inwards as well as the inward propagation of eccentricity. Then, the superhump behavior moves towards the state called stage B. (d) Finally, the disk becomes steady, and the temperature of the whole disk is higher than the temperature in state (a). 
        }
	\label{sed}
\end{figure}

The following works support our interpretation.  
The numerical simulation of the outbursts of X-ray 
transient systems show the mass accretion rate onto 
the central star rapidly rises after the onset of the tidal 
instability  \citep{ich94bxhn}.  
Also, the same kind of interpretation that the effect 
of tidal instability penetrates towards the innermost disk 
is proposed in dwarf-nova studies{from 
the softening of X-ray spectra confirmed in 
a well-observed WZ Sge-type DN just after the stage A to B 
transition of superhumps \citep{neu18j1222}.
The delay of superhump appearance is considered to depend on 
the binary mass ratio and the accumulated mass in the disk 
before each outburst \citep{osa02wzsgehump}.  
Our proposed interpretation could also explain the variety 
in the timing of the X-ray spectral transition including 
the standard hard-to-soft transition.  
This study therefore proposes one of the possible scenarios 
of a long standing question in high-energy astrophysics,
``what the trigger of the dramatic X-ray spectral transition 
is'' in transient low-mass X-ray binaries \citep{dun08gx339,shi19j1820}.
 
\subsubsection*{Acknowledgements}
We acknowledge the variable star observations from the 
AAVSO International Database contributed by observers 
worldwide and used in this research.
This work was financially supported by Grants-in-Aid 
for JSPS Fellows for young researchers (M.~Kimura and 
K.~Isogai).  

\subsubsection*{Author Contributions}
   K.N. and Y.W. led the campaign.  
   K.N. performed optical data analysis and compiled all optical data.
   M.K., N.K., and K.I. performed optical data analysis.
   M.K., T.K., D.N., and M.S. contributed to science discussions.
   M.K. performed multi-wavelength data analysis.
   Other authors than those mentioned above performed optical 
   observations.
   K.N., M.K., and Y.W. wrote the manuscript.
   T.K. supervised this project.
   K.N., M.K., T.K., W.Y., M.S., J.U., C.G., D.D., M.R., R.P., F.H.,
   E.M. and C.L. improved the manuscript.
   All authors have read and approved the manuscript.

\subsubsection*{Competing Interests}
The authors declare that they have no competing 
financial interests.

\subsubsection*{Author Information}
Correspondence and requests for materials should be addressed to 
K.N. \\ (niijima@kusastro.kyoto-u.ac.jp).

\newpage

\section*{METHODS}

\section{Detailed methods of optical observations and analyses}

After the onset of the 2018 outburst of ASASSN-18ey, 
the VSNET collaboration team \citep{VSNET} started a worldwide 
photometric campaign.  
Time-resolved CCD photometry was carried out at 32 sites 
using 32 telescopes with apertures of dozens of centimeters.
We also used the public AAVSO data\footnote{http://www.aavso.org/data-download/}.
We corrected for bias and flat-fielding in the usual manner,
performed standard aperture photometry, and measured 
magnitudes of ASASSN-18ey relative to local comparison 
stars whose magnitudes were measured by A.~Henden (sequence 
15167RN) from the AAVSO Variable Star Database \citep{APASS}\footnote{http://www.aavso.org/}. 
We applied small zero-point corrections to some observers' 
measurements.
The typical exposure time was 50 seconds.
All of the observation times were converted to Barycentric 
Julian Date (BJD).

We performed the period analyses with the phase dispersion 
minimization method (PDM)  \citep{PDM}.  In the analyses, 
the global trend of the light curves is subtracted by 
locally weighted polynomial regression (LOWESS: 
 \citet{LOWESS}).  The 1$\sigma$ errors are calculated 
via the method in  \citep{fer89error,Pdot2}.  
The robustness of the PDM results is confirmed by 
a variety of bootstraps.  We made 100 samples, each containing
50\% of the observations chosen at random, 
and performed the PDM analyses for each sample.  
The bootstrap results correspond to the 90\% 
confidence intervals of $\theta$ statistics.  

The $O-C$ diagram in Figure 2 shows the time difference 
of the superhump periods on each superhump cycle, and 
the linear and non-linear slopes represent the constant 
and varying periods, respectively.  
The positive linear slope before 
BJD 2,458,291.7 in ASASSN-18ey means that the period was 
constant and slightly longer than the assumed period
during stage A (see also the top panel of Fig.~2a).  
The period suddenly dropped at BJD 2,458,291.7 and 
gradually increases during stage B.
We first made a template profile from superhumps 
during BJD 2.458,279--2,458,405 (corresponds to 2018 June 9--2018 October 13)
by folding them with 
the period estimated by PDM, and fitted superhumps 
at each cycle with the template to estimate the time 
of the superhump maxima ($O$).  
After that, the $O-C$ is calculated by subtracting 
the assumed fix period ($C$) $\times$ the superhump cycle 
($E$) from $O \times E$.  
The resultant times are given in Extended Data Table 2.

\section{Overall optical behavior}

We show all of our optical photometric data during
the main outburst in Figure 5.  
The optical light curves show a complex behavior.  
They show a rapid rise at the onset of the outburst and 
a steep decay just after that.  Then, radio flares 
were detected and actually sub-second optical flaring 
that is likely related to synchrotron emission was also 
discovered \citep{utt18atel11423,gan18atel11437,tet18atel11440,rus18atel11533,tru18atel11539}.  
These nonthermal components would affect the optical light 
curves when stage A superhumps began to develop.  
The fast optical variability induced by synchrotron emission 
was observed commonly in the hard state among 
some of black-hole transients \citep{gan08gx339,alf18v404cyg}.
Actually the optical flux was scattered with respect to 
the X-ray flux especially in the hard state \citep{shi19j1820}, 
which cannot be predicted from the correlation between 
the optical and X-ray flux originating from the irradiated 
accretion disk \citep{rus06OIRandXrayCorr}.  

The optical luminosity increased again since several days
before the transition from stage A to stage B.
Around that time, the radio activity was quenching,
which was coincident with the hard-to-soft X-ray
spectral transition \citep{tet18atel11831,bro18atel11887}.  
The optical rise at the onset of the superhumps may be consistent
with the sudden increase of the amount of angular momentum loss
due to the tidal instability, and seem to resemble the light variations
reproduced by numerical simulations in the past \citep{ich94bxhn}.

\begin{figure}[htbp]
	\centering
	\begin{minipage}[c]{1\hsize}
		\centering\includegraphics[width=120mm]{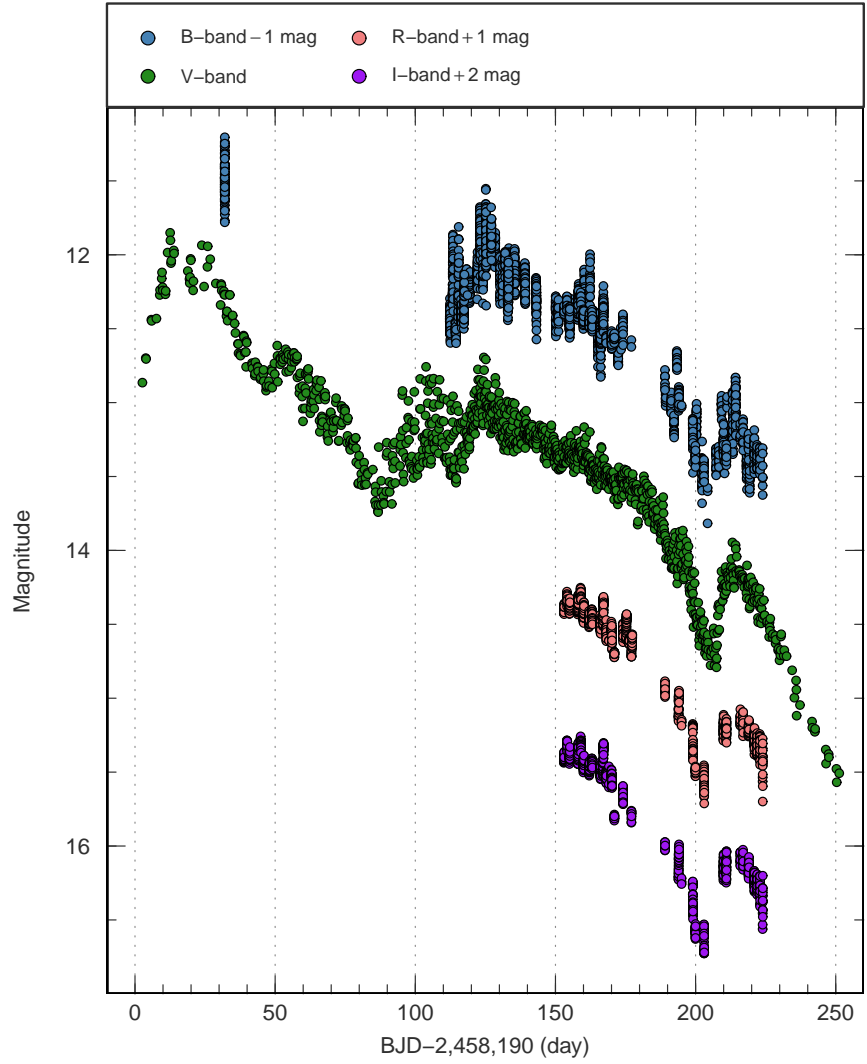}
	\end{minipage}
	\caption{\textbf{Multi-color light curves of ASASSN-18ey.}  The blue, green, pink, and purple points represent the light curves of $B$-band - 1mag, $V$-band, $R$-band + 1mag, and $I$-band + 2mag, respectively. The horizontal and vertical axis represent BJD-2,458,190 and Magnitude, respectively.}
        \label{lc}
\end{figure}

\section{SU UMa/WZ Sge-type stars and time evolution of their superhumps}

Dwarf novae (DNe) are a subtype of cataclysmic variables 
(CVs) and are semi-detached binaries composed of a white 
dwarf (the primary) and a late-type main-sequence star 
(the secondary).  There is an accretion disk around 
the primary.  Because of the thermal-viscous instability in the disk, 
episodic abrupt increases of luminosity occur, which 
are called ``outbursts'' \citep{war95book,osa96review}.

SU UMa-type DNe are a subclass of DNe
and show superoutbursts in addition to normal outbursts
with amplitudes of 2--5 mag and several-days durations.
Superoutbursts have $\sim$6-mag large amplitudes and
$\sim$2-weeks long durations, and moreover, small-amplitude
variations called ``superhumps'' with slightly longer periods
than that of orbital humps.
Superoutbursts and superhumps are believed to be 
the representation of the tidal instability \citep{whi88tidal,osa89suuma}.
Superhumps can be classified into three stages 
as to variations of periods and amplitudes: 
stage A, stage B, and stage C.  
During stage A, the period is constant and the amplitude 
increases with time.  During stage B, the period is 
varying and the amplitude gradually decreases.  
During stage C, the period is constant but shorter than 
the one during stage A, and the amplitude increases.

WZ Sge-type stars are an extreme subclass of DNe and 
belong to SU UMa-type DNe.  Their mass accretion rate 
is extremely small, and hence, the frequency of outbursts 
is very low.  The intervals between outbursts are typically 
larger than 5 years.  The two main observational features 
characterizing WZ Sge-type stars are early superhumps and 
rebrightenings.  
Early superhumps are observed immediately after 
the superoutburst and their period is almost equal to 
the orbital one.  Rebrightenings are sudden flux increases 
observed just after superoutbursts.

The period variation of superhumps is known to depend on
the mass ratio of binary systems.  
Figure 6 shows the $O-C$ diagram of superhumps 
in three systems, ASASSN-16dt, WZ Sge, and SW UMa.  
Their mass ratios are 0.036(2), 0.078(3), and 0.100(3), 
respectively.  
The smaller the mass ratio is, the longer 
the stage A duration is and the shorter the stage C duration 
is \citep{kat15wzsge}.  

The superhumps are excited by the strong tidal dissipation 
at the 3:1 resonance \citep{whi88tidal}.  The accretion disk 
becomes elliptical, and then, the eccentric accretion disk 
precesses \citep{hir90SHexcess}.  
The stage A superhumps are considered to represent the pure 
precession of the outermost ring of the accretion disk.  
The eccentricity wave propagates inwards with time and 
the pressure effect becomes significant.  When the superhumps 
enter stage B and the superhump period is not determined 
only by the dynamical precession, since the pressure effect is not 
negligible \citep{kat13qfromstageA,lub92SH,hir93SHperiod,mon01SH}.

\begin{figure}[htbp]
	\centering
	\begin{minipage}[c]{0.32\hsize}
		\centering\includegraphics[width=60mm]{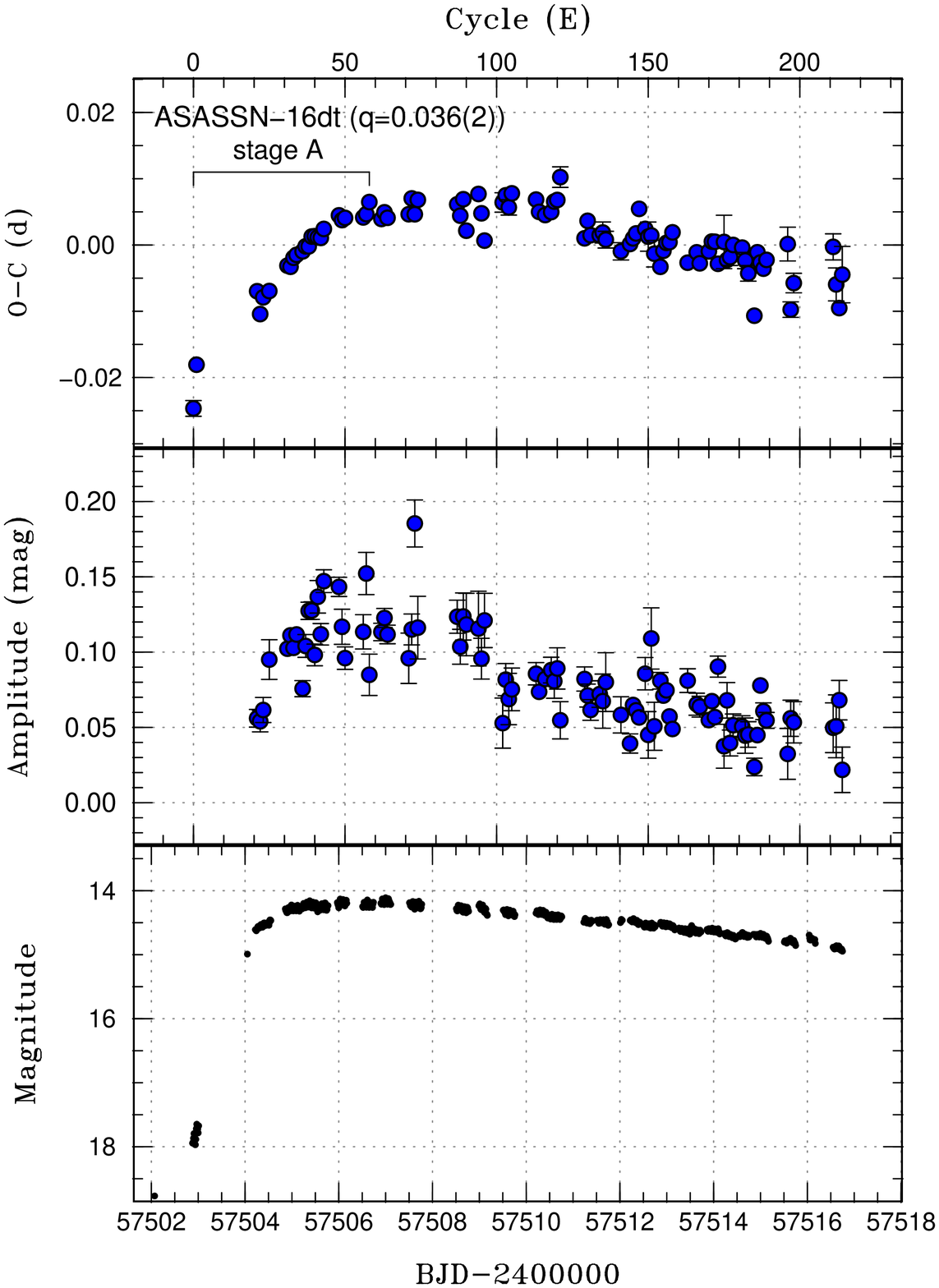}
	\end{minipage}
	\begin{minipage}[c]{0.32\hsize}
		\centering\includegraphics[width=60mm]{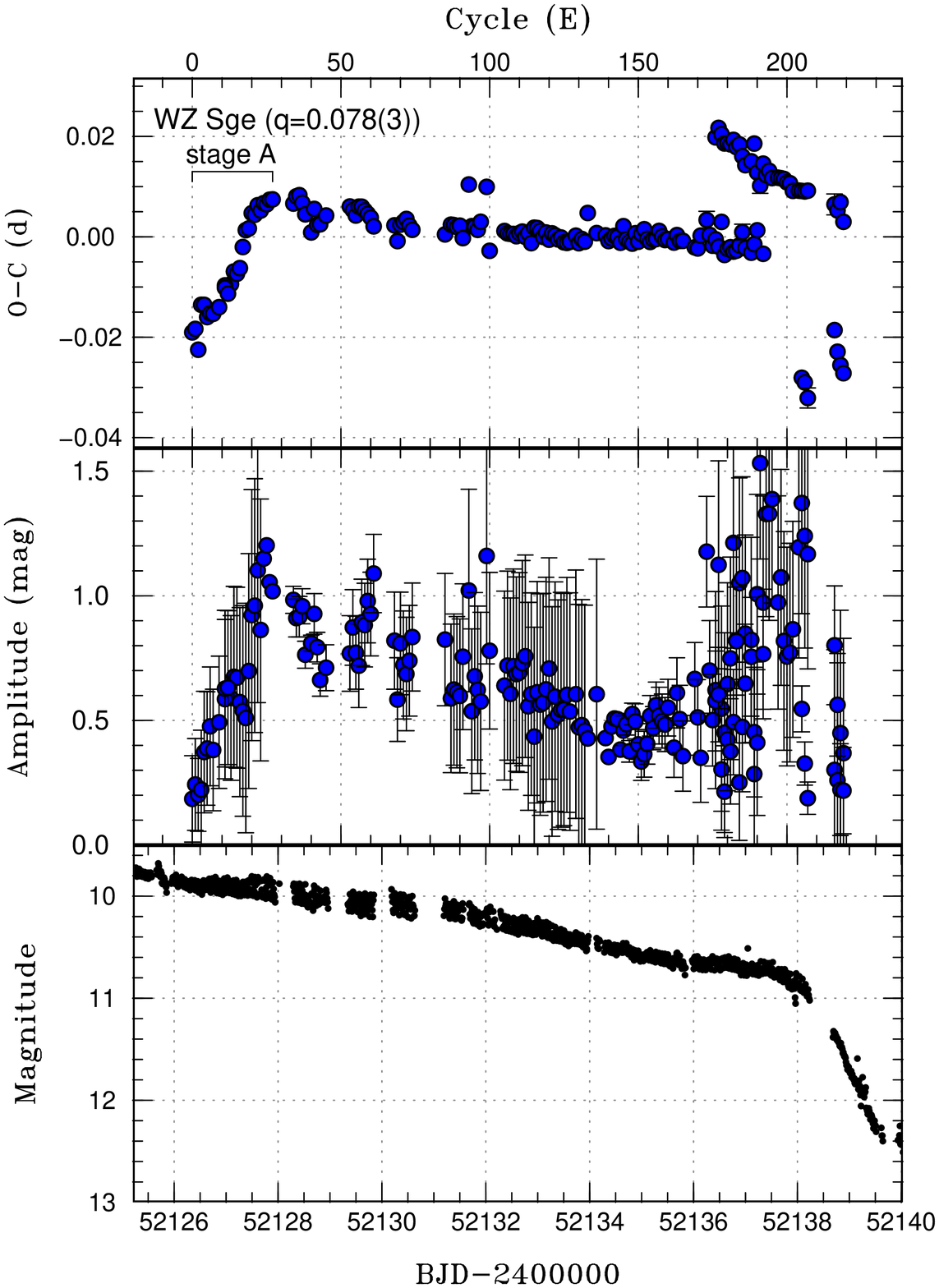}
	\end{minipage}
	\begin{minipage}[c]{0.32\hsize}
		\centering\includegraphics[width=60mm]{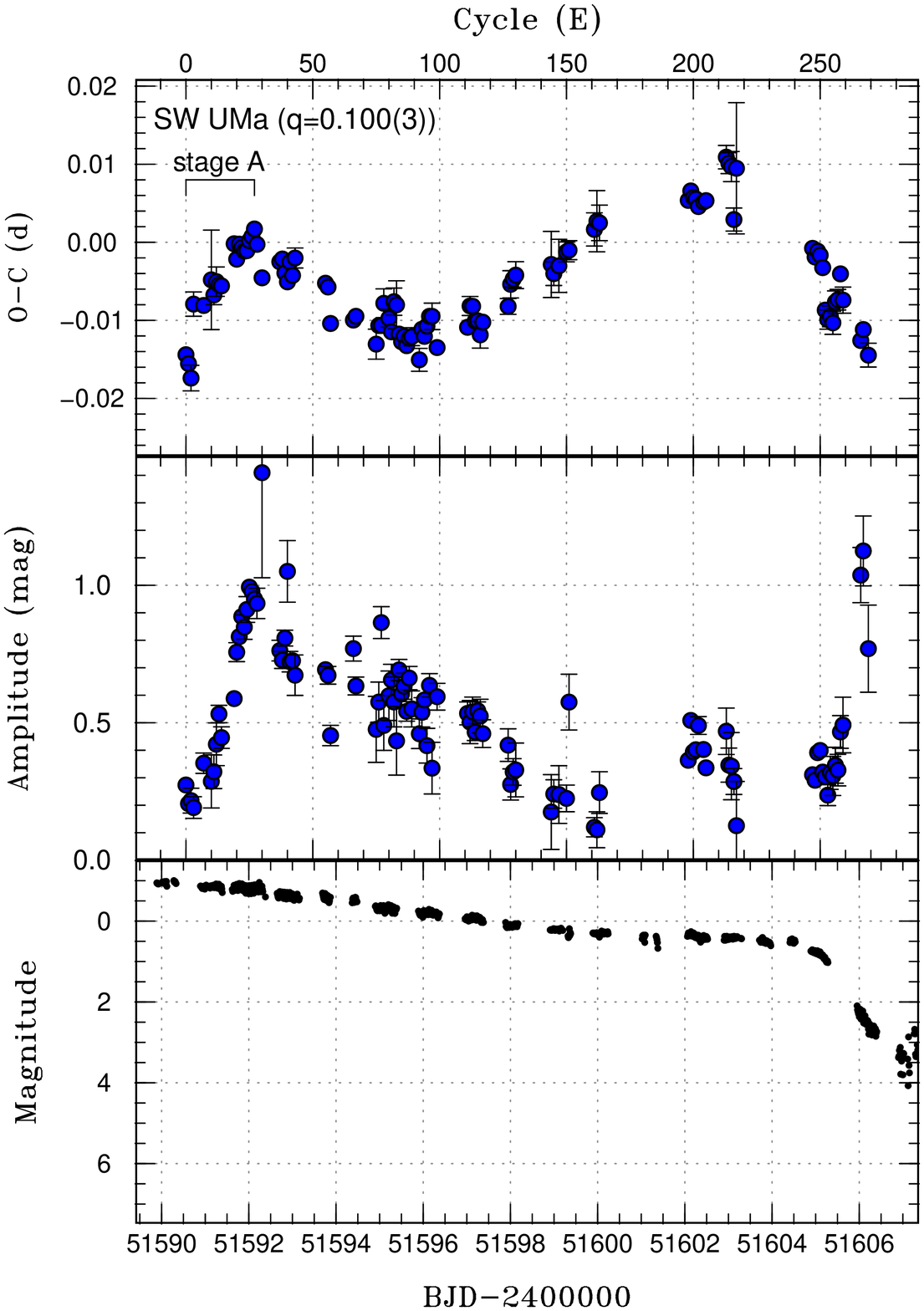}
	\end{minipage}
	\caption{\textbf{Typical period and amplitude variations of SU UMa-type stars during outburst.}  $O-C$ curves (upper panel), amplitudes of superhumps (middle panel) and light curves (lower panel) of SW UMa, WZ Sge and ASASSN-16dt. Each horizontal axis in units of BJD and cycle number is common to each of all column panels. The error bars represent 1$\sigma$ statistic errors.}
	\label{oc}
\end{figure}

\section{Estimation of black-hole masses with stage A superhumps}

The method for estimating the binary mass ratio from 
the stage A superhump period and the orbital period has been
developed in the study of SU UMa Sge-type stars.  
At the growing stage of superhumps (stage A), the superhumps 
are considered to represent the dynamical precession at the 
3:1 resonance radius \citep{osa13v1504cygKepler}.  
Here the growth time of superhumps is defined as the time 
until the eccentricity fully develops only at the 3:1 
resonance radius.
The fractional superhump excess $\varepsilon^{\ast}$ 
is defined as, 
$\varepsilon^{\ast} \equiv 1-P_{\rm orb}/P_{\rm SH}$,
and $\varepsilon^{\ast}$ of stage A superhumps 
is expressed by considering the dynamical precession as follows: 
\begin{equation}
\varepsilon^{\ast} = \frac{q}{\sqrt{1+q}} \left[\frac{1}{4} \frac{1}{\sqrt{r}} b_{3/2}^{(1)} \right], 
\label{omega_pr}
\end{equation}
where $r$ is the dimensionless radius normalized by 
the binary separation.  Here, $b_{3/2}^{(1)}$ is 
the Laplace coefficient \citep{hir90SHexcess,kat13qfromstageA}.  
By substituting the 3:1 resonance radius, which is defined 
as $3^{(-2/3)} (1+q)^{-1/3}$, to equation (\ref{omega_pr}), 
$\varepsilon^{\ast}$ depends only on the binary mass ratio $q$ 
 \citep{kat13qfromstageA}.  We can estimate the binary mass ratio 
only by measurements of the orbital and stage A superhump periods.

In addition, we estimated the binary mass ratio in ASASSN-18ey 
only with the duration of stage A superhumps.
Since superhumps are growing during stage A, the duration of 
stage A represents the growth time of the superhumps.
In addition, it was predicted theoretically that
the growth time of superhumps depends on $q^{-2}$ \citep{lub91SHb}.
The relation between $q$ and the duration of 
stage A for WZ Sge-type objects was confirmed by using 
the observational data and the above method and Figure 3 
shows the result \citep{kat13qfromstageA}.  
This relation is independent of the nature of 
the primary star, and hence, applicable for 
black-hole binaries.

The mass-ratio estimation by the empirical relation 
between the stage B superhumps and the superhump 
excess \citep{pat01SH,pat05SH} 
have used for a long time in many works including 
the very recent work \citep{tor19j1820}.  
However, this method has some uncertainty because 
the periods of the stage B superhumps are 
variable as shown in Figure 6.  
Also, it is revealed that this method easily underestimate 
the binary mass ratio \citep{nak13j2112j2037}.  
The method thus not suitable for the accurate estimation 
of black-hole masses.

\section{Spectral energy distribution modelling}

Figure 7 shows the multi-wavelength SEDs 
on July 8th (BJD 2,458,359) and August 19th (BJD 2,458,307) 
in 2018 when the source was simultaneously observed 
in the X-ray, ultraviolet (UV) and optical bands.  
The X-ray spectrum is extracted from simultaneous 
\textit{Swift}/XRT data (ObsIDs 00010627076 and 00088657009) 
which were taken in the WT mode.  The data are processed 
through the pipeline processing tool \texttt{xrtpipeline}.  
We select 1.0--10.0 keV XRT data.  
The UV flux is obtained from the Swift/UVOT images in 
the same ObsID as those of XRT through the standard tool 
\texttt{uvot2pha} provided by the Swift team.  
The optical flux is averaged per date.

\begin{figure}[htbp]
	\centering
	\begin{minipage}[c]{0.49\hsize}
		\centering\includegraphics[width=90mm]{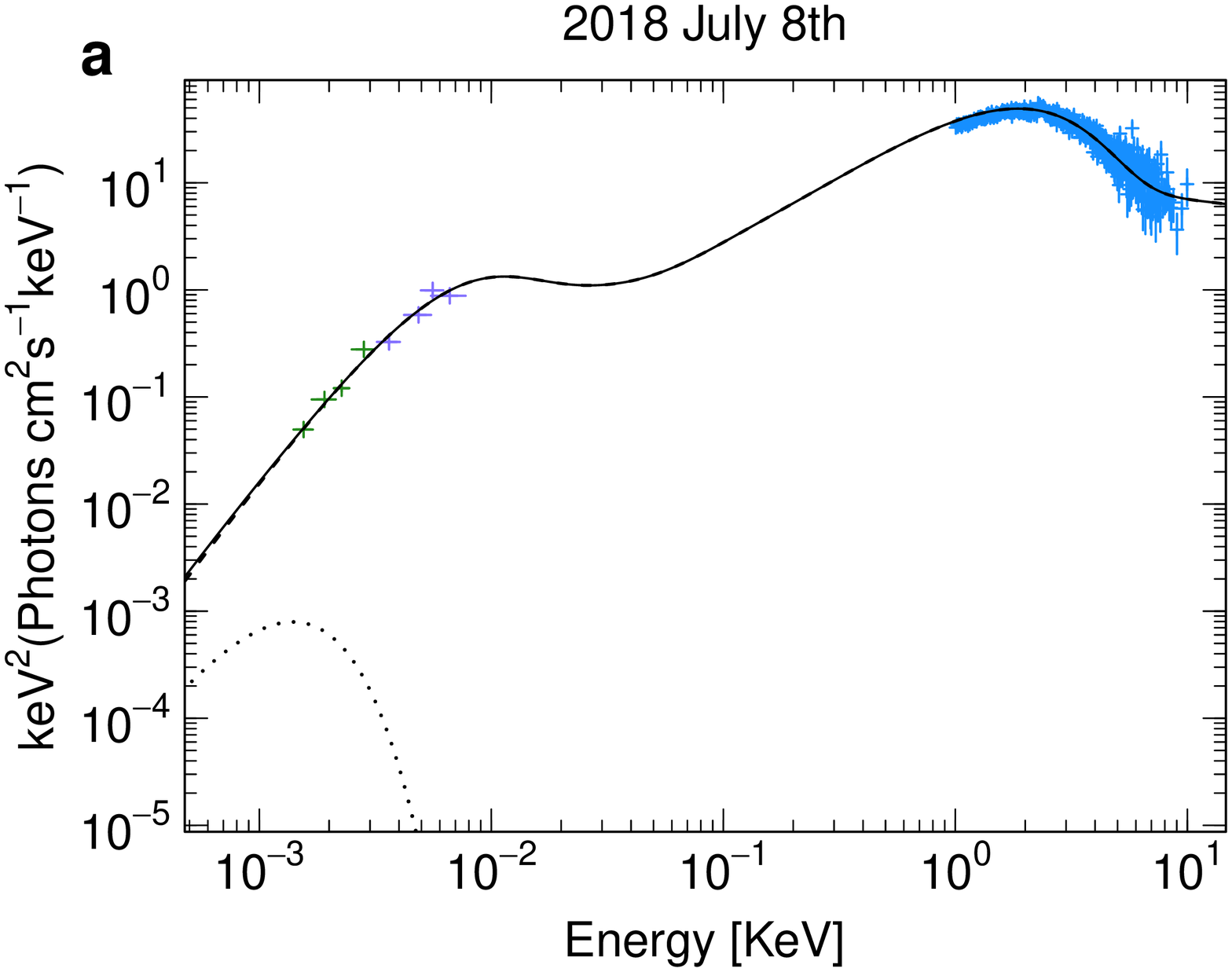}
	\end{minipage}
	\begin{minipage}[c]{0.49\hsize}
		\centering\includegraphics[width=90mm]{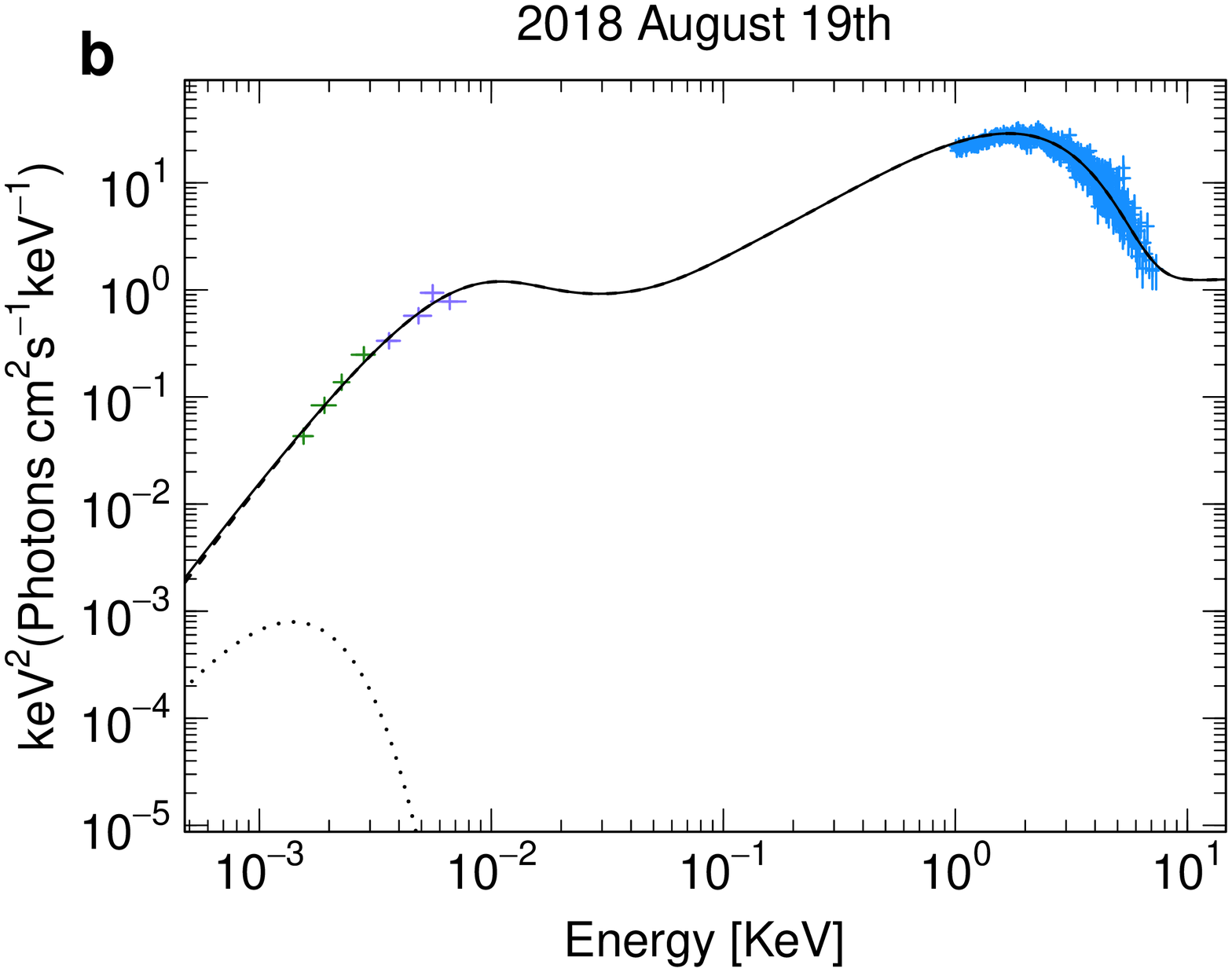}
	\end{minipage}
	\caption{\textbf{Simultaneous, 
	extinction-corrected multi-wavelength SEDs of ASASSN-18ey.}
    The broad-band spectra on (a) BJD 2,458,307 (corresponds to 2018 July 7)
    and (b) BJD 2,458,349 (corresponds to 2018 August 18).
    The blue, purple, and green crosses show 
    \textit{Swift} XRT, \textit{Swift} UVOT, and our optical data.  
    The optical fluxes are averaged over the intervals, and 
    the error bars represent their standard errors.
    The optical and $UV$ flux are deredden, and the X-ray flux 
    are deabsorbed.  
    The errors of X-ray and $UV$ data represent 1$\sigma$ 
    confidence intervals.  
    The dashed and dotted lines show the contribution of 
    emissions from the irradiated disk with Comptonisation and 
    from the companion star, respectively. }
	\label{sed}
\end{figure}

We fit the multi-wavelength spectra with the model 
\texttt{phabs*redden*(optxrplir+bbodyrad)}.  
The model \texttt{optxrplir} can simultaneously deal with 
the emission from an irradiated standard accretion disk and 
Comptonization component at the inner region.  The hot corona 
which emits Comptonized photons is assumed to be located 
around the central object, and it extends to $r_{\rm corona}$.  
The standard disk exists outside the hot corona and 
its outer part is reprocessed by X-rays emitted from 
the inner part of the disk and the corona.  
The models \texttt{phabs} and \texttt{redden} are implemented 
to correct interstellar absorption and extinction.  
The $N_H$ is fixed to 1.5$\times$10$^{21}$ 
cm$^{-2}$~ \citep{utt18atel11423}, and $E(B-V)$ is estimated to 
be 0.29 in fitting both of the two spectra.
With the model \texttt{bbodyrad}, we express the maximum 
contribution of the donor star to the optical flux, although 
it is less than 1\% compared with the contribution from 
the outer accretion disk.  We assume a K-type
main-sequence star as a donor star with a temperature of
4,130 K, and that the donor star fills its Roche-lobe \citep{all73quantities}.  
According to Gaia data release 2, the distance to ASASSN-18ey 
is $3.06^{+1.54}_{-0.82}$ kpc \citep{gan19gaiadistance}.  
In addition, we fix $kT_{\rm pl}$ in 
\texttt{optxrplir} to 100 keV, since it is difficult to 
constrain the parameter without hard X-ray data.  
Although \texttt{optxrplir} can treat low-temperature 
Comptonization component, we do not add it.  
The black-hole mass is estimated to be 6.7 $M_{\odot}$ in 
fitting simultaneously the two spectra, and we fix it to 
that value when fitting each spectrum separately.  
Then we assume the inclination angle is 60 deg.  

The best-fit values of $\log (L/L_{\rm Edd})$ and $r_{\rm corona}$ 
in units of $R_{\rm g}$, $\log (R_{\rm out}/R_{\rm g})$, and 
$f_{\rm out}$ are $-$0.70, 13, 5.1, and 0.088 as for ObsID 
00010627076, and $-$0.95, 7, 5.1, and 0.14 as for ObsID 
00088657009, respectively.  
Here, $L_{\rm Edd}$ is the Eddington luminosity, and 
$R_{\rm g}$ is the gravitational radius, respectively.  
The estimated disk radius is slightly less than the 3:1 
resonance radius, and appropriate as the disk radius after 
the onset of stage B superhumps \citep{ich93SHmasstransferburst}.  
The estimated $f_{\rm out}$ can be expected from the theoretical 
model of canonical irradiation on June 8th.  Then the irradiated 
flux is more than 3--5 larger than the underlying disk flux
expected from the standard temperature distribution.
This means X-ray irradiation governs the emission 
from the outer disk. 

The small-amplitude superhumps can be reproduced 
only by tidal dissipation as described in the main text.  
Although it is considered that the superhumps should not be 
reproduced only by the intrinsic luminosity of the viscous 
disk because of huge X-ray irradiation in black-hole 
binaries \citep{has01BHXNSH}, the actual irradiated flux 
in ASASSN-18ey was much smaller than expected in that study.  
Actually, the 14.9 $V$-band magnitude of the viscous disk 
on BJD 2,458,307 (corresponds to 2018 July 7) is
comparable to the initial magnitude 
on the date of the discovery of this outburst, which means 
the temperature at the outer part of the underlying viscous 
disk is high enough to keep the hot state by itself.  
Although the origin of the superhump amplitudes in ASASSN-18ey,
which are much larger than those of the past superhump candidates
observed in other black-hole binaries \citep{odo96BHXNSH,uem02j1118} is unclear, 
the hot surface area that is excited by tidal dissipation would 
be much larger because of the large disk expected from its long 
superhump period (see also Table 1), and 
the change in the vertical structure at the outer disk could 
be amplified via X-ray irradiation 
since X-ray irradiation makes the outer disk rim 
flared up \citep{cun76irradiation}.

\section{Variations of color indices}

We investigated the variations of color indices during 
BJD 2,458,302--2,458,310 (corresponds to 2018 July 2--2018 July 10), about a week after the onset of 
stage B superhumps.  
We estimated the $B-V$ and $V-R$ colors by approximating 
the simultaneous $V$-band magnitude to the $B$ and $R$ magnitudes 
with linear interpolation and averaged them with each of 
the stage A and stage B superhumps.  
The results are shown in Figure 8.
Although a slight color change is observable as for dwarf novae 
during stage B superhumps \citep{neu17j1222}, no appreciable 
color variations were detected in ASASSN-18ey during 
this time interval.  
The origin of color variations in dwarf novae is considered 
to be due to the enhanced viscous heating at the outer disk, 
and therefore, the color variation would not appear in black-hole 
binaries due to strong X-ray irradiation (Figure 7).  
Actually, our SED analyses suggest strong irradiation near 
this time period.  
In addition, no color variations in superhump candidates were 
confirmed also in GS 1124$-$68, another black-hole 
binary \citep{bai92gumus}.

\begin{figure}[htbp]
	\centering
	\begin{minipage}[c]{0.49\hsize}
		\centering\includegraphics[width=90mm]{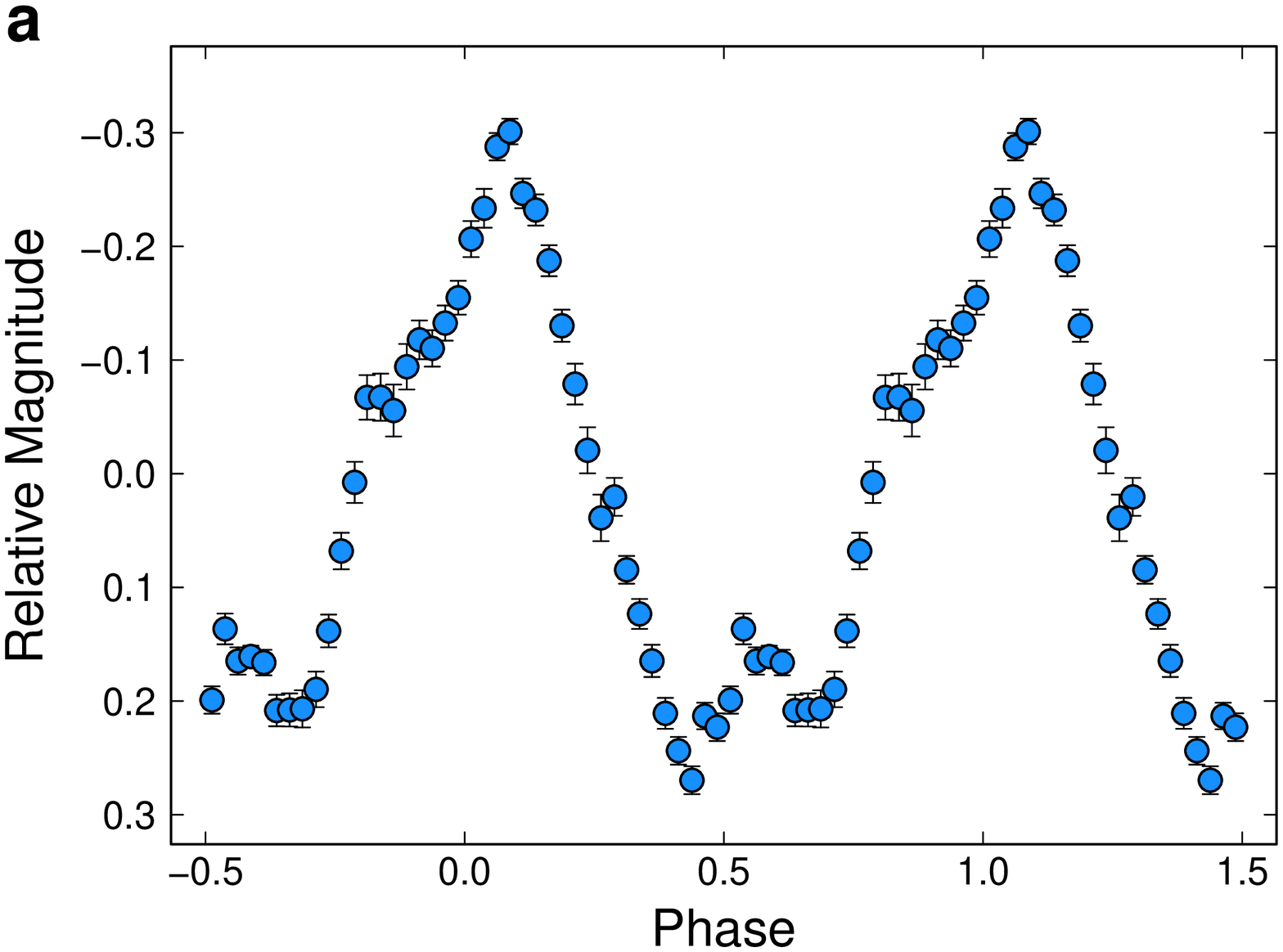}
	\end{minipage}
	\begin{minipage}[c]{0.49\hsize}
		\centering\includegraphics[width=90mm]{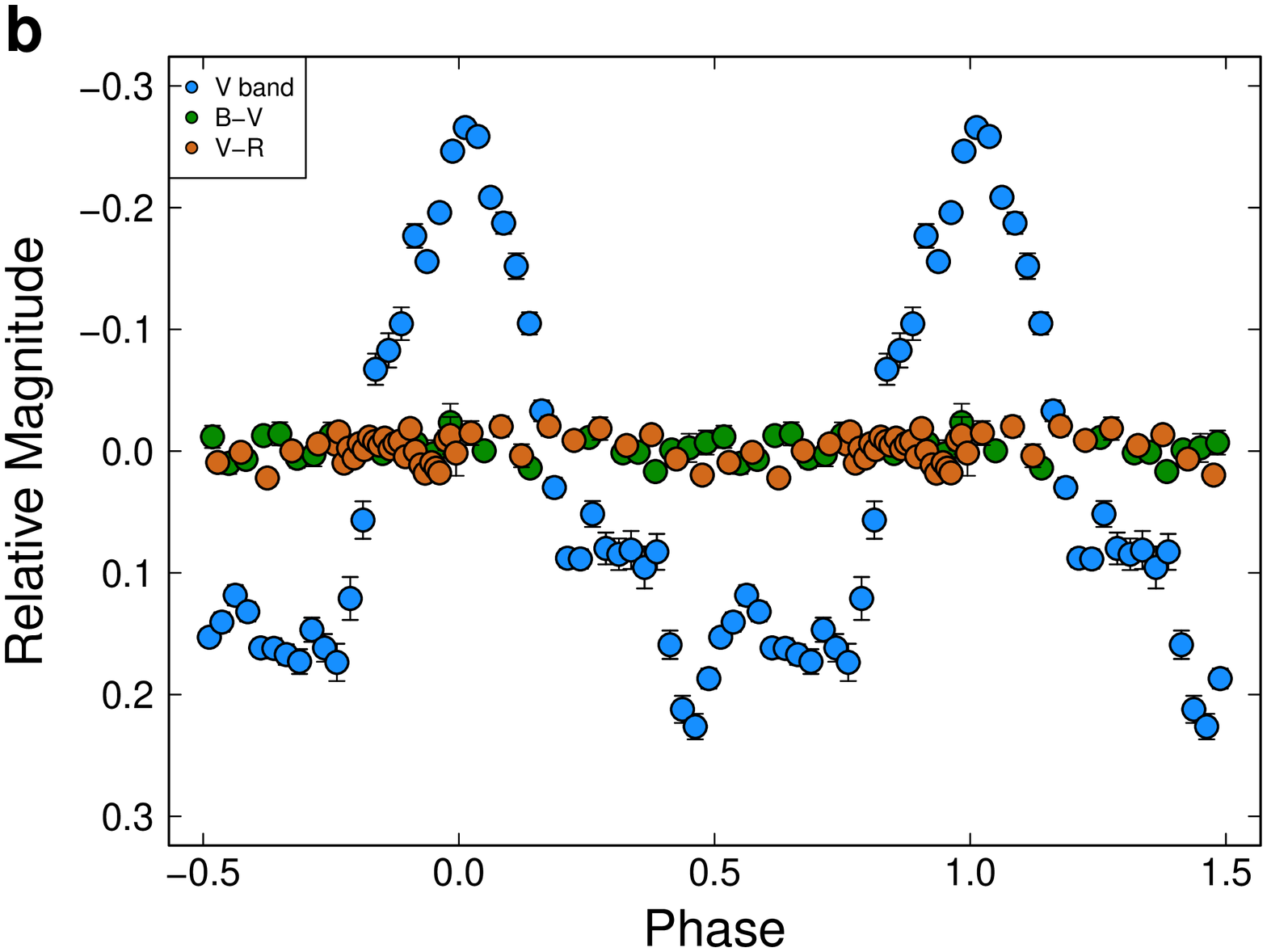}
	\end{minipage}
	\caption{\textbf{Phase-averaged profiles of superhumps.}  (a) Stage A superhump profile during BJD 2,458,279--2,458,295, and (b) stage B superhump profile during BJD 2,458,302--2,458,310.  As for stage B superhumps, we focus on the time interval when the multi-color light curves are obtained in order to plot the $B-V$ and $V-R$ color variations together.  The light curves are folded with the averaged period of each stage after subtracting the global trend of the light variations.  The error bars represent 1$\sigma$ statistic errors.}
	\label{phave}
\end{figure}

\begin{table*}[htb]
\caption{Objects Showing Superhumps during Outburst in Black-Hole Binaries.}
\label{tab:listsuperhump}
\begin{center}
\scalebox{0.9}{
\begin{tabular}{cccccc}
\hline
Object Name & Orbital Period [d] & Superhump Period [d] & $M_1$ [$M_\odot$] & Donor$^{*}$ & $q=M_2/M_1$ \\ \hline
GS 2000$+$25 & 0.344098(5)$^{[1]}$ & 0.3474(3)$^{[2]}$ & 8.5(1.5)$^{[3]}$ & K3--6/5$^{[4]}$ & 0.03--0.25$^{[3]}$ \\
(QZ Vul) & & & & & \\
GS 1124$-$68 & 0.4333(6)$^{[5]}$ & 0.4376(10)$^{[2]}$ & 6.95(60)$^{[6]}$ & K3--4/5$^{[7]}$ & 0.09--0.17$^{[6]}$ \\
(Nova Musca) & & & & & \\
GRO J0422$-$32 & 0.2121600(2)$^{[8]}$ & 0.2157(12)$^{[2]}$ & 3.57(34)$^{[9]}$ & M0--2/5$^{[9]}$ & 0.1093(86)$^{[9]}$ \\
(V518 Per) & & & & & \\
XTE J1118$+$480 & 0.169930(4)$^{[10, 11]}$ & 0.170529(6)$^{[12]}$ & 6.0--7.7$^{[11]}$ & K5/5$^{[13]}$ & 0.05$^{[14]}$ \\
(KV UMa) & & & & & \\
GRS 1716$-$249 & -- & 0.6127$^{[15]}$ & $>$4.9$^{[15]}$ & M0-5/5$^{[16]}$ & -- \\
(V2293 Oph) & & & & & \\
GRS 1009$-$45 & 0.285206(14)$^{[17]}$ & 0.1996$^{[18]}$ & 4.4$^{[17]}$ & K7-M0/5$^{[17]}$ & 7.3(0.8)$^{[17]}$ \\
(MM Vel) & & & & & \\
XTE J1859$+$226 & 0.274(2)$^{[19]}$ & 0.38385(73)$^{[20]}$ & $>$5.4$^{[19]}$ & G5$^{[21]}$ & -- \\
(V406 Vul) & & & & & \\
MAXI J1820$+$070 & -- & 0.688907(9) & -- & -- & $\sim$0.1 \\
(ASASSN-18ey) & & & & & \\
\hline
\multicolumn{6}{l}{$^{*}$ Spectral type.}\\
\multicolumn{6}{l}{\parbox{500pt}{
References: [1] \citet{che93qzvul}, [2] \citet{odo96BHXNSH}, [3] \citet{oro03BHB}, [4] \citet{har96qzvul}, [5] \citet{rem92gumus}, [6] \citet{gel01gumus}, [7] \citet{cas97gumus}, [8] \citet{web00v518perTiO}, [9] \citet{gel03v518per}, [10] \citet{wag01j1118}, [11] \citet{mcc01j1118mass}, [12] \citet{uem02j1118} , [13] \citet{gel06kvuma}, [14] \citet{gon14a0620kvuma}, [15] \citet{mas96grs1716SH}, [16] \citet{cha02IRobsBH}, [17] \citet{fil99mmvel}, [18] \citet{mas97mmvel}, [19] \citet{cor11j1859}, [20] \citet{uem04j1859}, [21] \citet{fil01j1859}
}} \\
\end{tabular}
}
\end{center}
\end{table*}

\section{Black-hole binaries showing superhump candidates in the past}

There are several transient black-hole binaries that 
have shown superhump candidates during outburst.
However, the observational quality is not enough to 
surely confirm whether they are superhumps and 
the period variation characteristic to superhumps 
observed in SU UMa stars were not detected.  
We discuss each object in which superhump candidates were 
observed during outburst as follows and list these objects 
and their properties in Table 1.  

The first suggestion that transient black-hole binaries
may exhibit superhumps similar to those seen in SU UMa stars
was made after the 1988 outburst in GS 2000+25 (QZ Vul) \citep{cha91qzvul}. 
They argued that the observed hump profile was not caused
by ellipsoidal variations.   
Additionally, X-ray heating of the donor star was ruled out
because no color variations were observed in these humps.  
Moreover, geometrical effects of the outer accretion disk 
were also excluded because of the expected low inclination.  
However, the number of data points were less than several hundred,
and they did not demonstrate they are clearly superhumps \citep{cha91qzvul}.
After that work, the orbital period was measured by
other authors \citep{che93qzvul}, which proved the estimated superhump period was 
slightly longer than the orbital period \citep{odo96BHXNSH}.  

In the 1991 outburst of GS 1124$-$68, similar humps to 
those observed in GS 2000$+$25 with a period slightly longer
than the orbital period were found \citep{bai92gumus}, 
and then, they were suggested to be superhumps as discussed 
in GS 2000$+$25 \citep{cha91qzvul}.  However, the coverage 
of the data is sparse and the total length of the observations 
was only a few days during the outburst which typically 
continues for more than 100 days.  
The data were reanalyzed after their work, and 
the period of these humps were estimated to be 
shorter \citep{odo96BHXNSH}.  Since the data quality is thus
not at all convincing, it is difficult to distinguish
the superhump period that they argued from the orbital period 
measured by other authors \citep{rem92gumus}.  

Two types of optical variability were discovered 
in GRO J0422$+$32 (V518 Per), another black-hole binary, 
during its 1992--1993 outburst.  The observed dip-type 
variations with a shorter period and another kind of 
variations with a longer period were considered to be 
the orbital modulations and superhumps, respectively \citep{kat95v518per}.  
The data used in this paper had been the densest one 
among the data taken in the above three systems prior 
to our work, and they succeeded to estimate the superhump period 
within a 0.5\% accuracy.  
They noticed the hump period changed with time during 
outburst, but did not confirm if this feature originated 
from the actual period change of superhumps, and interpreted 
the period decrease as being due to the appearance of
the orbital modulations.
Soon after that work, the hump periods at the late stage 
of the outburst and in quiescence were estimated 
by  \citep{che95v518per}, and they were identical with each other.  
The authors hence regard them as orbital modulations.  
Also, the periods were comparable to the hump period 
at the latter stage of the outburst in  \citep{kat95v518per}.   

Several years later, XTE J1118$+$480 (KV UMa), a newly 
discovered X-ray transient, showed during its 2000 outburst 
optical modulations having periods slightly longer than 
the orbital period \citep{uem02j1118,wag01j1118,mcc01j1118spec}.  
The hump period was estimated within the 0.05\% accuracy, 
and thus, they clearly detected the existence of superhumps 
during outburst for the first time by using their extensive data. 
Since the hump profile was not easily explained by the orbital 
periods, the detected continuous period decreases strongly suggested
the variations are superhumps.
In this object, the period variations of superhump candidates 
were detected for the first time among black-hole binaries.  
However, the authors were not able to classify the superhumps into 
each stage, so the certain detection of humps quite similar 
to those observed in SU UMa stars was not completed.  
This is because they did not obtain the data of the initial 
stage of superhumps, and because the detailed definition of 
the stages of superhumps was established in 2009 \citep{Pdot} 
even in SU UMa stars. 
Also, this system is not classical because it did not enter 
the canonical spectral state evolution of outbursts 
in comparison with many other black-hole transients 
 \citep{tan96XNreview,che97BHXN}.  
Therefore our study is the first one showing the absolute 
discovery of superhumps in normal transient black-hole 
X-ray binaries.  
In addition, the superhump modulations were also confirmed 
in the rebrightenings in 2005 \citep{zur06j1118}.  

Although the observational data were sparse due to 
the dimness of the outbursts, there are three other objects 
possibly showing superhumps during outburst.  
It is unclear because of their unknown orbital periods, but 
it was argued that superhump-like modulations were found 
in GRS 1716$-$249 and GRS 1009$-$45 during 
outburst \citep{mas96grs1716SH,mas97mmvel}.  
Additionally, coherent optical variations appeared in 
XTE J1859$+$226, which have possibly a longer period 
than the orbital one \citep{zur02j1859,uem04j1859}, 
though the period is much longer than the orbital period 
lately measured by the optical photometry of ellipsoidal 
modulations in quiescence \citep{cor11j1859}.  

As described above, it is not rare that superhumps 
are observed during outburst in black-hole binaries.  
The most plausible model of the origin of superhumps is 
the precession of the eccentric disk exerted by the tidal 
instability, which is triggered in small mass-ratio binary 
systems.  The condition is considered to be $q \lesssim 
0.25$ \citep{whi88tidal,osa89suuma,osa02wzsgehump}.  
Generally, most of the black-hole low-mass X-ray binaries would
satisfy this condition. However, the optical outbursts in this kind of 
objects are usually dim due to their distance, and hence, 
this may have prevented universal detections of superhumps 
in black hole X-ray binaries.
Nevertheless, the current development of optical telescopes 
and observational networks will enable us to detect superhumps 
in many transient black-hole binaries.

\newcommand{\noop}[1]{}

\newpage

\section*{Extended Data tables}
\setcounter{table}{0}
\renewcommand{\thetable}{E.\arabic{table}}

\begin{table*}[htb]
\caption{Log of observations of the 2018 outburst in ASASSN-18ey.}
\label{log}
\begin{center}
\begin{tabular}{rrrrrc}
\hline
${\rm Start}^{*}$ & ${\rm End}^{*}$ & ${\rm Mag}^{\dagger}$ & ${\rm Error}^{\ddagger}$ & $N^{\S}$ & ${\rm Obs}^{\parallel}$ \\ \hline  
58192.8343 & 58192.9583 & 12.945 & 0.002 & 3230 & LCO \\
58193.8548 & 58193.9223 & 12.761 & 0.002 & 1758 & LCO \\
58195.7754 & 58195.8155 & 12.565 & 0.009 & 206 & HaC \\
58195.8259 & 58195.9040 & 12.500 & 0.002 & 1895 & LCO \\
58196.8272 & 58196.8776 & 12.539 & 0.007 & 261 & HaC \\
58197.5629 & 58197.6761 & 12.513 & 0.005 & 293 & BSM \\
58197.8165 & 58197.8803 & 12.471 & 0.004 & 320 & HaC \\
\hline
\end{tabular}
\end{center}
\end{table*}

\setcounter{table}{0}
\begin{table*}[htb]
\caption{Log of observations of the 2018 outburst in ASASSN-18ey (continued).}
\label{log}
\begin{center}
\begin{tabular}{rrrrrc}
\hline
${\rm Start}^{*}$ & ${\rm End}^{*}$ & ${\rm Mag}^{\dagger}$ & ${\rm Error}^{\ddagger}$ & $N^{\S}$ & ${\rm Obs}^{\parallel}$ \\ \hline
58208.9486 & 58209.0466 & 12.083 & 0.007 & 134 & SGE \\
58209.4513 & 58209.6169 & 12.088 & 0.004 & 188 & OAR \\
58209.7823 & 58209.8812 & 12.251 & 0.003 & 837 & HaC \\
58209.8194 & 58209.9373 & 12.078 & 0.005 & 269 & UJH \\
58209.8862 & 58209.9326 & 12.256 & 0.007 & 90 & BJA \\
58209.9487 & 58210.0212 & 12.158 & 0.005 & 250 & SGE \\
58210.7795 & 58210.8811 & 12.298 & 0.003 & 875 & HaC \\
\hline
\end{tabular}
\end{center}
\end{table*}

\setcounter{table}{0}
\begin{table*}[htb]
\caption{Log of observations of the 2018 outburst in ASASSN-18ey (continued).}
\label{log}
\begin{center}
\begin{tabular}{rrrrrc}
\hline
${\rm Start}^{*}$ & ${\rm End}^{*}$ & ${\rm Mag}^{\dagger}$ & ${\rm Error}^{\ddagger}$ & $N^{\S}$ & ${\rm Obs}^{\parallel}$ \\ \hline
58221.7839 & 58221.8867 & 12.268 & 0.002 & 1212 & HaC \\
58221.9131 & 58221.9545 & 12.582 & 0.005 & 253 & COO \\
58221.9501 & 58221.9707 & 12.618 & 0.007 & 196 & LPA \\
58221.9721 & 58221.9954 & 11.759 & 0.011 & 199 & LPA \\
58221.9972 & 58222.0187 & 12.250 & 0.007 & 198 & LPA \\
58222.0201 & 58222.0316 & 12.451 & 0.015 & 71 & LPA \\
58222.5303 & 58222.6026 & 12.717 & 0.002 & 4035 & Trt \\
\hline
\end{tabular}
\end{center}
\end{table*}

\setcounter{table}{0}
\begin{table*}[htb]
\caption{Log of observations of the 2018 outburst in ASASSN-18ey (continued).}
\label{log}
\begin{center}
\begin{tabular}{rrrrrc}
\hline
${\rm Start}^{*}$ & ${\rm End}^{*}$ & ${\rm Mag}^{\dagger}$ & ${\rm Error}^{\ddagger}$ & $N^{\S}$ & ${\rm Obs}^{\parallel}$ \\ \hline
58237.8006 & 58237.9131 & 13.068 & 0.003 & 326 & BJA \\
58237.8609 & 58237.9618 & 12.864 & 0.003 & 424 & SGE \\
58238.7551 & 58238.8896 & 12.789 & 0.006 & 193 & HaC \\
58238.7768 & 58238.8752 & 12.833 & 0.006 & 235 & UJH \\
58238.8098 & 58238.8739 & 12.747 & 0.004 & 254 & RIT \\
58239.4609 & 58239.6014 & 12.845 & 0.003 & 505 & BSM \\
58239.7352 & 58239.8867 & 12.786 & 0.006 & 109 & HaC \\
\hline
\end{tabular}
\end{center}
\end{table*}

\setcounter{table}{0}
\begin{table*}[htb]
\caption{Log of observations of the 2018 outburst in ASASSN-18ey (continued).}
\label{log}
\begin{center}
\begin{tabular}{rrrrrc}
\hline
${\rm Start}^{*}$ & ${\rm End}^{*}$ & ${\rm Mag}^{\dagger}$ & ${\rm Error}^{\ddagger}$ & $N^{\S}$ & ${\rm Obs}^{\parallel}$ \\ \hline
58249.5178 & 58249.6025 & 12.879 & 0.002 & 215 & RJW \\
58249.6250 & 58249.8895 & 12.975 & 0.007 & 195 & HaC \\
58249.7702 & 58249.7899 & 12.915 & 0.010 & 67 & LPA \\
58249.9000 & 58249.9792 & 13.055 & 0.002 & 213 & SGE \\
58250.4588 & 58250.5633 & 13.011 & 0.002 & 1985 & Trt \\
58250.6223 & 58250.8924 & 12.906 & 0.008 & 199 & HaC \\
58250.8439 & 58250.9865 & 12.783 & 0.003 & 47 & SGE \\
\hline
\end{tabular}
\end{center}
\end{table*}

\setcounter{table}{0}
\begin{table*}[htb]
\caption{Log of observations of the 2018 outburst in ASASSN-18ey (continued).}
\label{log}
\begin{center}
\begin{tabular}{rrrrrc}
\hline
${\rm Start}^{*}$ & ${\rm End}^{*}$ & ${\rm Mag}^{\dagger}$ & ${\rm Error}^{\ddagger}$ & $N^{\S}$ & ${\rm Obs}^{\parallel}$ \\ \hline
58259.5984 & 58259.8923 & 13.089 & 0.005 & 174 & HaC \\
58259.6505 & 58259.6988 & 13.053 & 0.004 & 74 & LMA \\
58259.6748 & 58259.8198 & 12.994 & 0.002 & 745 & RIT \\
58260.4809 & 58260.6567 & 13.357 & 0.004 & 130 & Van \\
58260.6704 & 58260.8911 & 13.144 & 0.004 & 119 & HaC \\
58260.8292 & 58260.9710 & 13.033 & 0.002 & 353 & SGE \\
58261.4276 & 58261.6702 & 13.272 & 0.005 & 177 & Van \\
\hline
\end{tabular}
\end{center}
\end{table*}

\setcounter{table}{0}
\begin{table*}[htb]
\caption{Log of observations of the 2018 outburst in ASASSN-18ey (continued).}
\label{log}
\begin{center}
\begin{tabular}{rrrrrc}
\hline
${\rm Start}^{*}$ & ${\rm End}^{*}$ & ${\rm Mag}^{\dagger}$ & ${\rm Error}^{\ddagger}$ & $N^{\S}$ & ${\rm Obs}^{\parallel}$ \\ \hline
58272.9310 & 58272.9835 & 13.340 & 0.004 & 140 & SGE \\
58272.9782 & 58272.9983 & 13.489 & 0.010 & 32 & LPA \\
58273.3974 & 58273.6108 & 13.514 & 0.004 & 485 & BSM \\
58273.4839 & 58273.5822 & 13.778 & 0.004 & 244 & Van \\
58273.8636 & 58273.9845 & 13.551 & 0.003 & 340 & SGE \\
58274.4821 & 58274.5353 & 14.065 & 0.003 & 97 & Van \\
58274.8055 & 58274.8676 & 13.455 & 0.011 & 39 & HaC \\
\hline
\end{tabular}
\end{center}
\end{table*}

\setcounter{table}{0}
\begin{table*}[htb]
\caption{Log of observations of the 2018 outburst in ASASSN-18ey (continued).}
\label{log}
\begin{center}
\begin{tabular}{rrrrrc}
\hline
${\rm Start}^{*}$ & ${\rm End}^{*}$ & ${\rm Mag}^{\dagger}$ & ${\rm Error}^{\ddagger}$ & $N^{\S}$ & ${\rm Obs}^{\parallel}$ \\ \hline
58285.5803 & 58285.8652 & 13.482 & 0.005 & 213 & HaC \\
58285.6117 & 58285.8497 & 13.479 & 0.002 & 567 & RIT \\
58285.7729 & 58285.7799 & 13.661 & 0.006 & 10 & DJE \\
58285.8151 & 58285.9554 & 13.468 & 0.003 & 400 & SGE \\
58286.3816 & 58286.4951 & 13.768 & 0.002 & 220 & Van \\
58286.4437 & 58286.5911 & 13.576 & 0.002 & 253 & BSM \\
58286.6710 & 58286.7967 & 13.256 & 0.004 & 207 & LMA \\
\hline
\end{tabular}
\end{center}
\end{table*}

\setcounter{table}{0}
\begin{table*}[htb]
\caption{Log of observations of the 2018 outburst in ASASSN-18ey (continued).}
\label{log}
\begin{center}
\begin{tabular}{rrrrrc}
\hline
${\rm Start}^{*}$ & ${\rm End}^{*}$ & ${\rm Mag}^{\dagger}$ & ${\rm Error}^{\ddagger}$ & $N^{\S}$ & ${\rm Obs}^{\parallel}$ \\ \hline
58293.3733 & 58293.6679 & 13.241 & 0.003 & 625 & ATE \\
58293.3817 & 58293.6264 & 13.248 & 0.003 & 302 & Van \\
58293.4165 & 58293.5333 & 13.261 & 0.001 & 608 & Trt \\
58293.5667 & 58293.8427 & 12.952 & 0.020 & 74 & HaC \\
58293.6426 & 58293.9268 & 12.968 & 0.005 & 672 & SJI \\
58293.8493 & 58293.9841 & 13.068 & 0.005 & 380 & SGE \\
58293.9673 & 58294.2454 & 13.354 & 0.003 & 613 & MGW \\
\hline
\end{tabular}
\end{center}
\end{table*}

\setcounter{table}{0}
\begin{table*}[htb]
\caption{Log of observations of the 2018 outburst in ASASSN-18ey (continued).}
\label{log}
\begin{center}
\begin{tabular}{rrrrrc}
\hline
${\rm Start}^{*}$ & ${\rm End}^{*}$ & ${\rm Mag}^{\dagger}$ & ${\rm Error}^{\ddagger}$ & $N^{\S}$ & ${\rm Obs}^{\parallel}$ \\ \hline
58300.0805 & 58300.1387 & 12.952 & 0.007 & 117 & Kis \\
58300.3395 & 58300.4454 & -0.422 & 0.002 & 318 & Lic \\
58300.3985 & 58300.5016 & 13.384 & 0.003 & 144 & EHE \\
58300.4088 & 58300.5205 & 13.342 & 0.007 & 77 & Van \\
58300.4119 & 58300.5156 & 13.375 & 0.003 & 193 & IMi \\
58300.4274 & 58300.5903 & 13.261 & 0.005 & 354 & RJW \\
58300.6086 & 58300.7397 & 12.909 & 0.005 & 280 & RIT \\
\hline
\end{tabular}
\end{center}
\end{table*}

\setcounter{table}{0}
\begin{table*}[htb]
\caption{Log of observations of the 2018 outburst in ASASSN-18ey (continued).}
\label{log}
\begin{center}
\begin{tabular}{rrrrrc}
\hline
${\rm Start}^{*}$ & ${\rm End}^{*}$ & ${\rm Mag}^{\dagger}$ & ${\rm Error}^{\ddagger}$ & $N^{\S}$ & ${\rm Obs}^{\parallel}$ \\ \hline
58304.3407 & 58304.4486 & -0.276 & 0.001 & 326 & Lic \\
58304.3726 & 58304.6202 & 13.549 & 0.002 & 306 & Van \\
58304.3844 & 58304.4992 & 13.528 & 0.002 & 153 & ATE \\
58304.5381 & 58304.8126 & 13.327 & 0.017 & 119 & HaC \\
58304.5982 & 58304.8476 & 13.227 & 0.007 & 504 & RIT \\
58304.6987 & 58304.7329 & 13.175 & 0.006 & 100 & SGE \\
58305.3188 & 58305.5180 & -0.721 & 0.025 & 38 & CRI \\
\hline
\end{tabular}
\end{center}
\end{table*}

\setcounter{table}{0}
\begin{table*}[htb]
\caption{Log of observations of the 2018 outburst in ASASSN-18ey (continued).}
\label{log}
\begin{center}
\begin{tabular}{rrrrrc}
\hline
${\rm Start}^{*}$ & ${\rm End}^{*}$ & ${\rm Mag}^{\dagger}$ & ${\rm Error}^{\ddagger}$ & $N^{\S}$ & ${\rm Obs}^{\parallel}$ \\ \hline
58308.2883 & 58308.3388 & -0.763 & 0.009 & 12 & CRI \\
58308.2891 & 58308.3450 & 0.353 & 0.076 & 16 & CRI \\
58308.2898 & 58308.3312 & 0.676 & 0.013 & 10 & CRI \\
58308.2913 & 58308.3373 & 1.380 & 0.011 & 11 & CRI \\
58308.3347 & 58308.4468 & -0.568 & 0.001 & 279 & Lic \\
58308.3672 & 58308.6206 & 13.246 & 0.001 & 375 & Van \\
58308.4161 & 58308.5953 & 13.257 & 0.001 & 458 & RJW \\
\hline
\end{tabular}
\end{center}
\end{table*}

\setcounter{table}{0}
\begin{table*}[htb]
\caption{Log of observations of the 2018 outburst in ASASSN-18ey (continued).}
\label{log}
\begin{center}
\begin{tabular}{rrrrrc}
\hline
${\rm Start}^{*}$ & ${\rm End}^{*}$ & ${\rm Mag}^{\dagger}$ & ${\rm Error}^{\ddagger}$ & $N^{\S}$ & ${\rm Obs}^{\parallel}$ \\ \hline
58311.8357 & 58311.8742 & 12.999 & 0.002 & 108 & SGE \\
58311.9275 & 58312.1961 & 13.125 & 0.002 & 533 & MGW \\
58312.0588 & 58312.2624 & 13.333 & 0.007 & 66 & Ioh \\
58312.0601 & 58312.2712 & 13.162 & 0.006 & 70 & Ioh \\
58312.3204 & 58312.5829 & 13.062 & 0.004 & 591 & BSM \\
58312.3267 & 58312.4381 & -1.005 & 0.005 & 69 & Lic \\
58312.3272 & 58312.4385 & -0.807 & 0.005 & 69 & Lic \\
\hline
\end{tabular}
\end{center}
\end{table*}

\setcounter{table}{0}
\begin{table*}[htb]
\caption{Log of observations of the 2018 outburst in ASASSN-18ey (continued).}
\label{log}
\begin{center}
\begin{tabular}{rrrrrc}
\hline
${\rm Start}^{*}$ & ${\rm End}^{*}$ & ${\rm Mag}^{\dagger}$ & ${\rm Error}^{\ddagger}$ & $N^{\S}$ & ${\rm Obs}^{\parallel}$ \\ \hline
58315.1365 & 58315.2740 & 12.881 & 0.012 & 97 & Ioh \\
58315.2497 & 58315.4167 & 13.044 & 0.002 & 200 & OYE \\
58315.3208 & 58315.4305 & -0.811 & 0.004 & 58 & Lic \\
58315.3213 & 58315.4310 & -0.607 & 0.004 & 58 & Lic \\
58315.3220 & 58315.4317 & -0.392 & 0.004 & 58 & Lic \\
58315.3405 & 58315.5083 & 13.040 & 0.002 & 464 & Trt \\
58315.3679 & 58315.4733 & 13.219 & 0.002 & 113 & GCH \\
\hline
\end{tabular}
\end{center}
\end{table*}

\setcounter{table}{0}
\begin{table*}[htb]
\caption{Log of observations of the 2018 outburst in ASASSN-18ey (continued).}
\label{log}
\begin{center}
\begin{tabular}{rrrrrc}
\hline
${\rm Start}^{*}$ & ${\rm End}^{*}$ & ${\rm Mag}^{\dagger}$ & ${\rm Error}^{\ddagger}$ & $N^{\S}$ & ${\rm Obs}^{\parallel}$ \\ \hline
58318.9048 & 58319.1232 & 13.165 & 0.001 & 482 & MGW \\
58318.9761 & 58319.1880 & 0.726 & 0.001 & 1569 & KU1 \\
58319.3010 & 58319.3925 & 0.616 & 0.003 & 204 & MLF \\
58319.3244 & 58319.5831 & 13.238 & 0.003 & 582 & BSM \\
58319.3695 & 58319.6290 & 13.138 & 0.003 & 293 & Van \\
58319.4039 & 58319.6034 & 13.428 & 0.002 & 616 & CDZ \\
58319.6731 & 58319.9020 & 13.024 & 0.003 & 552 & SGE \\
\hline
\end{tabular}
\end{center}
\end{table*}

\setcounter{table}{0}
\begin{table*}[htb]
\caption{Log of observations of the 2018 outburst in ASASSN-18ey (continued).}
\label{log}
\begin{center}
\begin{tabular}{rrrrrc}
\hline
${\rm Start}^{*}$ & ${\rm End}^{*}$ & ${\rm Mag}^{\dagger}$ & ${\rm Error}^{\ddagger}$ & $N^{\S}$ & ${\rm Obs}^{\parallel}$ \\ \hline
58323.4076 & 58323.5085 & 13.166 & 0.008 & 89 & EHE \\
58323.4248 & 58323.5925 & 13.524 & 0.003 & 577 & CDZ \\
58323.6709 & 58323.9001 & 13.201 & 0.002 & 643 & SGE \\
58323.7016 & 58323.9370 & 13.300 & 0.002 & 550 & JJI \\
58323.8838 & 58324.1634 & 13.106 & 0.003 & 617 & MGW \\
58324.1292 & 58324.2413 & 0.770 & 0.003 & 869 & KU1 \\
58324.3993 & 58324.5895 & 13.623 & 0.001 & 664 & CDZ \\
\hline
\end{tabular}
\end{center}
\end{table*}

\setcounter{table}{0}
\begin{table*}[htb]
\caption{Log of observations of the 2018 outburst in ASASSN-18ey (continued).}
\label{log}
\begin{center}
\begin{tabular}{rrrrrc}
\hline
${\rm Start}^{*}$ & ${\rm End}^{*}$ & ${\rm Mag}^{\dagger}$ & ${\rm Error}^{\ddagger}$ & $N^{\S}$ & ${\rm Obs}^{\parallel}$ \\ \hline
58330.3503 & 58330.5614 & 13.212 & 0.001 & 516 & ATE \\
58330.3933 & 58330.4937 & 13.491 & 0.001 & 348 & CDZ \\
58330.4111 & 58330.5073 & 13.529 & 0.001 & 155 & Van \\
58330.5481 & 58330.7424 & 13.243 & 0.003 & 92 & HaC \\
58330.6106 & 58330.7191 & 13.201 & 0.001 & 130 & LMA \\
58330.6858 & 58330.9110 & 13.588 & 0.001 & 608 & COO \\
58330.8628 & 58331.1443 & 13.207 & 0.001 & 836 & MGW \\
\hline
\end{tabular}
\end{center}
\end{table*}

\clearpage

\setcounter{table}{0}
\begin{table*}[htb]
\caption{Log of observations of the 2018 outburst in ASASSN-18ey (continued).}
\label{log}
\begin{center}
\begin{tabular}{rrrrrc}
\hline
${\rm Start}^{*}$ & ${\rm End}^{*}$ & ${\rm Mag}^{\dagger}$ & ${\rm Error}^{\ddagger}$ & $N^{\S}$ & ${\rm Obs}^{\parallel}$ \\ \hline
58336.2419 & 58336.4445 & 0.476 & 0.002 & 581 & MLF \\
58336.3569 & 58336.4992 & 13.568 & 0.001 & 234 & Van \\
58336.3762 & 58336.4530 & 13.202 & 0.002 & 218 & Trt \\
58336.3773 & 58336.5332 & 13.224 & 0.001 & 392 & RJW \\
58336.3873 & 58336.4033 & 13.266 & 0.002 & 41 & CDZ \\
58336.3999 & 58336.5037 & 13.239 & 0.001 & 134 & EHE \\
58336.5441 & 58336.7260 & 13.307 & 0.002 & 85 & HaC \\
\hline
\end{tabular}
\end{center}
\end{table*}

\setcounter{table}{0}
\begin{table*}[htb]
\caption{Log of observations of the 2018 outburst in ASASSN-18ey (continued).}
\label{log}
\begin{center}
\begin{tabular}{rrrrrc}
\hline
${\rm Start}^{*}$ & ${\rm End}^{*}$ & ${\rm Mag}^{\dagger}$ & ${\rm Error}^{\ddagger}$ & $N^{\S}$ & ${\rm Obs}^{\parallel}$ \\ \hline
58341.3006 & 58341.5024 & 13.419 & 0.002 & 167 & BSM \\
58341.3570 & 58341.4806 & 3.314 & 0.001 & 253 & Vih \\
58341.3934 & 58341.5305 & 13.375 & 0.001 & 124 & Van \\
58341.5212 & 58341.7103 & 13.429 & 0.003 & 80 & HaC \\
58341.5785 & 58341.7499 & 13.434 & 0.001 & 316 & DKS \\
58341.6579 & 58341.7842 & 13.449 & 0.002 & 201 & SGE \\
58342.2555 & 58342.4217 & 13.487 & 0.001 & 200 & OYE \\
\hline
\end{tabular}
\end{center}
\end{table*}

\setcounter{table}{0}
\begin{table*}[htb]
\caption{Log of observations of the 2018 outburst in ASASSN-18ey (continued).}
\label{log}
\begin{center}
\begin{tabular}{rrrrrc}
\hline
${\rm Start}^{*}$ & ${\rm End}^{*}$ & ${\rm Mag}^{\dagger}$ & ${\rm Error}^{\ddagger}$ & $N^{\S}$ & ${\rm Obs}^{\parallel}$ \\ \hline
58346.3459 & 58346.5001 & 13.779 & 0.003 & 270 & Van \\
58346.3459 & 58346.5001 & 13.783 & 0.003 & 270 & Van \\
58346.4090 & 58346.4872 & 13.495 & 0.003 & 103 & EHE \\
58346.5170 & 58346.6985 & 13.435 & 0.005 & 74 & HaC \\
58346.6022 & 58346.6963 & 13.525 & 0.004 & 115 & MZK \\
58346.6530 & 58346.8403 & 13.302 & 0.002 & 300 & SGE \\
58346.9407 & 58347.0485 & 13.461 & 0.004 & 122 & Nel \\
\hline
\end{tabular}
\end{center}
\end{table*}

\setcounter{table}{0}
\begin{table*}[htb]
\caption{Log of observations of the 2018 outburst in ASASSN-18ey (continued).}
\label{log}
\begin{center}
\begin{tabular}{rrrrrc}
\hline
${\rm Start}^{*}$ & ${\rm End}^{*}$ & ${\rm Mag}^{\dagger}$ & ${\rm Error}^{\ddagger}$ & $N^{\S}$ & ${\rm Obs}^{\parallel}$ \\ \hline
58350.3040 & 58350.4417 & 3.157 & 0.001 & 269 & Vih \\
58350.5228 & 58350.6880 & 13.391 & 0.003 & 75 & HaC \\
58350.6696 & 58350.7930 & 13.399 & 0.001 & 200 & SGE \\
58350.8859 & 58351.0870 & 13.300 & 0.003 & 385 & MGW \\
58351.2864 & 58351.4594 & 3.293 & 0.002 & 404 & Vih \\
58351.2885 & 58351.4158 & -0.491 & 0.008 & 28 & CRI \\
58351.2893 & 58351.4166 & 0.430 & 0.006 & 28 & CRI \\
\hline
\end{tabular}
\end{center}
\end{table*}

\setcounter{table}{0}
\begin{table*}[htb]
\caption{Log of observations of the 2018 outburst in ASASSN-18ey (continued).}
\label{log}
\begin{center}
\begin{tabular}{rrrrrc}
\hline
${\rm Start}^{*}$ & ${\rm End}^{*}$ & ${\rm Mag}^{\dagger}$ & ${\rm Error}^{\ddagger}$ & $N^{\S}$ & ${\rm Obs}^{\parallel}$ \\ \hline
58355.5087 & 58355.6734 & 13.456 & 0.008 & 75 & HaC \\
58355.6443 & 58355.7679 & 13.361 & 0.002 & 191 & SGE \\
58355.9467 & 58356.1758 & 1.494 & 0.003 & 68 & KU2 \\
58355.9474 & 58356.1765 & 1.090 & 0.004 & 65 & KU2 \\
58355.9481 & 58356.1743 & 0.669 & 0.004 & 65 & KU2 \\
58355.9488 & 58356.1483 & -0.048 & 0.005 & 51 & KU2 \\
58355.9697 & 58356.1699 & 13.574 & 0.004 & 142 & Ioh \\
\hline
\end{tabular}
\end{center}
\end{table*}

\setcounter{table}{0}
\begin{table*}[htb]
\caption{Log of observations of the 2018 outburst in ASASSN-18ey (continued).}
\label{log}
\begin{center}
\begin{tabular}{rrrrrc}
\hline
${\rm Start}^{*}$ & ${\rm End}^{*}$ & ${\rm Mag}^{\dagger}$ & ${\rm Error}^{\ddagger}$ & $N^{\S}$ & ${\rm Obs}^{\parallel}$ \\ \hline
58360.2819 & 58360.4577 & 3.474 & 0.001 & 410 & Vih \\
58360.3473 & 58360.5082 & 13.631 & 0.008 & 85 & Van \\
58360.3819 & 58360.4909 & 13.607 & 0.003 & 131 & CDZ \\
58360.4929 & 58360.6560 & 13.468 & 0.003 & 81 & HaC \\
58360.6770 & 58360.7705 & 13.521 & 0.002 & 140 & SGE \\
58360.9628 & 58361.1389 & 13.656 & 0.004 & 230 & Ioh \\
58360.9678 & 58361.0182 & 1.625 & 0.003 & 14 & KU2 \\
\hline
\end{tabular}
\end{center}
\end{table*}

\setcounter{table}{0}
\begin{table*}[htb]
\caption{Log of observations of the 2018 outburst in ASASSN-18ey (continued).}
\label{log}
\begin{center}
\begin{tabular}{rrrrrc}
\hline
${\rm Start}^{*}$ & ${\rm End}^{*}$ & ${\rm Mag}^{\dagger}$ & ${\rm Error}^{\ddagger}$ & $N^{\S}$ & ${\rm Obs}^{\parallel}$ \\ \hline
58364.6343 & 58364.7671 & 13.465 & 0.002 & 200 & SGE \\
58365.2238 & 58365.3853 & 13.923 & 0.008 & 53 & Zub \\
58365.2426 & 58365.3785 & 0.658 & 0.003 & 367 & MLF \\
58365.3487 & 58365.4769 & 13.487 & 0.002 & 315 & CDZ \\
58365.3563 & 58365.4683 & 13.511 & 0.003 & 197 & Van \\
58365.4938 & 58365.6459 & 13.533 & 0.003 & 77 & HaC \\
58365.5498 & 58365.5841 & 13.546 & 0.004 & 30 & LMA \\
\hline
\end{tabular}
\end{center}
\end{table*}

\setcounter{table}{0}
\begin{table*}[htb]
\caption{Log of observations of the 2018 outburst in ASASSN-18ey (continued).}
\label{log}
\begin{center}
\begin{tabular}{rrrrrc}
\hline
${\rm Start}^{*}$ & ${\rm End}^{*}$ & ${\rm Mag}^{\dagger}$ & ${\rm Error}^{\ddagger}$ & $N^{\S}$ & ${\rm Obs}^{\parallel}$ \\ \hline
58373.2964 & 58373.4255 & 13.806 & 0.004 & 83 & BSM \\
58373.3427 & 58373.4573 & 13.885 & 0.005 & 115 & Van \\
58373.4993 & 58373.6243 & 13.665 & 0.008 & 49 & HaC \\
58373.5495 & 58373.7137 & 13.642 & 0.004 & 126 & UJH \\
58374.2547 & 58374.4185 & 3.424 & 0.002 & 383 & Vih \\
58374.2984 & 58374.3904 & 13.595 & 0.003 & 61 & BSM \\
58374.3372 & 58374.4522 & 13.612 & 0.001 & 282 & CDZ \\
\hline
\end{tabular}
\end{center}
\end{table*}

\setcounter{table}{0}
\begin{table*}[htb]
\caption{Log of observations of the 2018 outburst in ASASSN-18ey (continued).}
\label{log}
\begin{center}
\begin{tabular}{rrrrrc}
\hline
${\rm Start}^{*}$ & ${\rm End}^{*}$ & ${\rm Mag}^{\dagger}$ & ${\rm Error}^{\ddagger}$ & $N^{\S}$ & ${\rm Obs}^{\parallel}$ \\ \hline
58381.2471 & 58381.4000 & 3.823 & 0.003 & 357 & Vih \\
58381.3274 & 58381.4306 & 14.105 & 0.004 & 213 & CDZ \\
58381.5007 & 58381.6024 & 14.077 & 0.005 & 33 & HaC \\
58381.5279 & 58381.6373 & 14.158 & 0.004 & 84 & RIT \\
58381.6186 & 58381.7067 & 14.117 & 0.002 & 150 & SGE \\
58381.8747 & 58382.0051 & 13.968 & 0.001 & 276 & MGW \\
58382.2331 & 58382.3327 & 1.258 & 0.005 & 267 & MLF \\
\hline
\end{tabular}
\end{center}
\end{table*}

\setcounter{table}{0}
\begin{table*}[htb]
\caption{Log of observations of the 2018 outburst in ASASSN-18ey (continued).}
\label{log}
\begin{center}
\begin{tabular}{rrrrrc}
\hline
${\rm Start}^{*}$ & ${\rm End}^{*}$ & ${\rm Mag}^{\dagger}$ & ${\rm Error}^{\ddagger}$ & $N^{\S}$ & ${\rm Obs}^{\parallel}$ \\ \hline
58385.6116 & 58385.7038 & 14.162 & 0.003 & 150 & SGE \\
58385.8748 & 58385.9115 & 14.162 & 0.004 & 82 & MGW \\
58386.2392 & 58386.2654 & 3.833 & 0.003 & 62 & Vih \\
58386.2935 & 58386.3964 & 14.135 & 0.002 & 220 & RJW \\
58386.3218 & 58386.4509 & 14.172 & 0.004 & 203 & Van \\
58386.3421 & 58386.4197 & 14.226 & 0.004 & 81 & CDZ \\
58386.4784 & 58386.5955 & 14.077 & 0.006 & 75 & LMA \\
\hline
\end{tabular}
\end{center}
\end{table*}

\setcounter{table}{0}
\begin{table*}[htb]
\caption{Log of observations of the 2018 outburst in ASASSN-18ey (continued).}
\label{log}
\begin{center}
\begin{tabular}{rrrrrc}
\hline
${\rm Start}^{*}$ & ${\rm End}^{*}$ & ${\rm Mag}^{\dagger}$ & ${\rm Error}^{\ddagger}$ & $N^{\S}$ & ${\rm Obs}^{\parallel}$ \\ \hline
58390.9710 & 58390.9806 & 14.339 & 0.007 & 22 & MGW \\
58391.2277 & 58391.3037 & 1.731 & 0.011 & 192 & MLF \\
58391.2510 & 58391.3760 & 14.640 & 0.007 & 57 & BSM \\
58391.2699 & 58391.3956 & 14.835 & 0.002 & 219 & Van \\
58391.2844 & 58391.4151 & 14.504 & 0.002 & 284 & RJW \\
58391.3152 & 58391.4252 & 14.575 & 0.002 & 205 & CDZ \\
58391.4709 & 58391.6241 & 14.453 & 0.005 & 97 & LMA \\
\hline
\end{tabular}
\end{center}
\end{table*}

\setcounter{table}{0}
\begin{table*}[htb]
\caption{Log of observations of the 2018 outburst in ASASSN-18ey (continued).}
\label{log}
\begin{center}
\begin{tabular}{rrrrrc}
\hline
${\rm Start}^{*}$ & ${\rm End}^{*}$ & ${\rm Mag}^{\dagger}$ & ${\rm Error}^{\ddagger}$ & $N^{\S}$ & ${\rm Obs}^{\parallel}$ \\ \hline
58397.5876 & 58397.7044 & 14.645 & 0.006 & 141 & SJI \\
58397.8928 & 58397.9730 & 14.509 & 0.006 & 49 & Nel \\
58397.9129 & 58398.0153 & 14.446 & 0.010 & 78 & Ioh \\
58397.9162 & 58397.9587 & 14.440 & 0.005 & 54 & MGW \\
58398.2232 & 58398.3349 & 0.716 & 0.009 & 19 & CRI \\
58398.2244 & 58398.3360 & 1.552 & 0.009 & 19 & CRI \\
58398.2255 & 58398.3372 & 1.985 & 0.012 & 19 & CRI \\
\hline
\end{tabular}
\end{center}
\end{table*}

\setcounter{table}{0}
\begin{table*}[htb]
\caption{Log of observations of the 2018 outburst in ASASSN-18ey (continued).}
\label{log}
\begin{center}
\begin{tabular}{rrrrrc}
\hline
${\rm Start}^{*}$ & ${\rm End}^{*}$ & ${\rm Mag}^{\dagger}$ & ${\rm Error}^{\ddagger}$ & $N^{\S}$ & ${\rm Obs}^{\parallel}$ \\ \hline
58403.4747 & 58403.5418 & 14.188 & 0.004 & 46 & HaC \\
58403.5630 & 58403.6256 & 14.254 & 0.007 & 42 & UJH \\
58403.6399 & 58403.7222 & 14.250 & 0.003 & 95 & SGE \\
58404.2374 & 58404.3359 & 0.428 & 0.017 & 22 & CRI \\
58404.2374 & 58404.3359 & 0.428 & 0.017 & 22 & Vih \\
58404.2382 & 58404.3367 & 1.155 & 0.008 & 22 & CRI \\
58404.2390 & 58404.3329 & 1.509 & 0.008 & 18 & CRI \\
\hline
\end{tabular}
\end{center}
\end{table*}

\setcounter{table}{0}
\begin{table*}[htb]
\caption{Log of observations of the 2018 outburst in ASASSN-18ey (continued).}
\label{log}
\begin{center}
\begin{tabular}{rrrrrc}
\hline
${\rm Start}^{*}$ & ${\rm End}^{*}$ & ${\rm Mag}^{\dagger}$ & ${\rm Error}^{\ddagger}$ & $N^{\S}$ & ${\rm Obs}^{\parallel}$ \\ \hline
58410.8711 & 58410.9710 & 0.823 & 0.014 & 29 & KU2 \\
58410.8734 & 58411.0177 & 1.473 & 0.006 & 46 & KU2 \\
58410.8779 & 58411.0192 & 2.041 & 0.005 & 48 & KU2 \\
58410.8786 & 58411.0319 & 1.793 & 0.005 & 49 & KU2 \\
58411.4799 & 58411.5186 & 14.257 & 0.009 & 18 & HaC \\
58411.5902 & 58411.6700 & 14.258 & 0.005 & 50 & SGE \\
58411.9043 & 58412.0146 & 2.069 & 0.006 & 36 & KU2 \\
\hline
\end{tabular}
\end{center}
\end{table*}

\setcounter{table}{0}
\begin{table*}[htb]
\caption{Log of observations of the 2018 outburst in ASASSN-18ey (continued).}
\label{log}
\begin{center}
\begin{tabular}{rrrrrc}
\hline
${\rm Start}^{*}$ & ${\rm End}^{*}$ & ${\rm Mag}^{\dagger}$ & ${\rm Error}^{\ddagger}$ & $N^{\S}$ & ${\rm Obs}^{\parallel}$ \\ \hline
58420.8670 & 58420.9873 & 0.221 & 0.004 & 163 & KU3 \\
58421.4828 & 58421.4923 & 14.714 & 0.016 & 17 & HaC \\
58421.6064 & 58421.6214 & 14.677 & 0.011 & 15 & SGE \\
58421.8985 & 58422.0082 & 0.238 & 0.005 & 148 & KU3 \\
58422.3238 & 58422.3568 & 14.936 & 0.008 & 37 & Van \\
58422.4830 & 58422.4895 & 14.719 & 0.026 & 12 & HaC \\
58422.8811 & 58422.9101 & 0.296 & 0.010 & 24 & KU3 \\
\end{tabular}
\end{center}
\end{table*}

\begin{table*}[htb]
\caption{Times of superhump maxima in ASASSN-18ey.  }
\label{max}
\begin{center}
\begin{tabular}{rllrr}
\hline
$E^{*}$ & ${\rm Max}^{\dagger}$ & Error & ${O - C}^{\ddagger}$ & $N^{\S}$  \\ \hline
0 & 58279.8440 & 0.0033 & -0.0870 & 731 \\
1 & 58280.4646 & 0.0009 & -0.1553 & 683 \\
2 & 58281.1999 & 0.0074 & -0.1089 & 489 \\
3 & 58281.8491 & 0.0016 & -0.1486 & 917 \\
4 & 58282.5680 & 0.0010 & -0.1186 & 623 \\
5 & 58283.3621 & 0.0009 & -0.0134 & 1123 \\
7 & 58284.6721 & 0.0009 & -0.0812 & 1882 \\
\hline
\end{tabular}
\end{center}
\end{table*}

\setcounter{table}{0}
\begin{table*}[htb]
\caption{Times of superhump maxima in ASASSN-18ey (continued).  }
\label{max}
\begin{center}
\begin{tabular}{rllrr}
\hline
$E^{*}$ & ${\rm Max}^{\dagger}$ & Error & ${O - C}^{\ddagger}$ & $N^{\S}$  \\ \hline
57 & 58319.2593 & 0.0009 & 0.0606 & 3039 \\
58 & 58319.8660 & 0.0005 & -0.0216 & 2394 \\
59 & 58320.5493 & 0.0012 & -0.0272 & 1144 \\
60 & 58321.2636 & 0.0008 & -0.0018 & 1927 \\
61 & 58321.9246 & 0.0006 & -0.0298 & 2198 \\
62 & 58322.6273 & 0.0005 & -0.0159 & 2725 \\
63 & 58323.3158 & 0.0019 & -0.0164 & 2081 \\
\hline
\end{tabular}
\end{center}
\end{table*}

\setcounter{table}{0}
\begin{table*}[htb]
\caption{Times of superhump maxima in ASASSN-18ey (continued).  }
\label{max}
\begin{center}
\begin{tabular}{rllrr}
\hline
$E^{*}$ & ${\rm Max}^{\dagger}$ & Error & ${O - C}^{\ddagger}$ & $N^{\S}$  \\ \hline
118 & 58361.2027 & 0.0032 & -0.0193 & 1511 \\
120 & 58362.5710 & 0.0016 & -0.0288 & 1043 \\
121 & 58363.2287 & 0.0072 & -0.0600 & 932 \\
122 & 58363.9226 & 0.0019 & -0.0550 & 750 \\
123 & 58364.6181 & 0.0026 & -0.0485 & 946 \\
124 & 58365.3381 & 0.0017 & -0.0174 & 1022 \\
125 & 58366.0460 & 0.0028 & 0.0016 & 335 \\
\end{tabular}
\end{center}
\end{table*}

\end{document}